\documentclass{jpp}
\usepackage{graphicx}

\usepackage[utf8]{inputenc}
\usepackage[T1]{fontenc}
\usepackage{amsmath}

\usepackage{graphicx}
\usepackage{bm}
\usepackage{physics}
\usepackage[colorlinks=true, linkcolor=black, citecolor=blue]{hyperref}
\usepackage[capitalise, nameinlink]{cleveref}

\shorttitle{Parker Spiral Plasmoids}
\shortauthor{E. E. Peterson et al.}

\title{Laminar and Turbulent Plasmoid Ejection in a Laboratory Parker Spiral Current Sheet}

\author{Ethan E. Peterson\aff{1}\corresp{\email{peterson@psfc.mit.edu}}, Douglass A. Endrizzi\aff{2}, Michael Clark\aff{2}, Jan Egedal\aff{2}, Kenneth Flanagan\aff{2}, Nuno F. Loureiro\aff{1}, Jason Milhone\aff{2}, Joseph Olson\aff{2}, Carl R. Sovinec\aff{3}, John Wallace\aff{2}, and Cary B. Forest\aff{2}
}
\affiliation{\aff{1}Plasma Science and Fusion Center, MIT, Cambridge, MA 02139, USA
\aff{2}Department of Physics, University of Wisconsin--Madison, Madison, WI 53706, USA \aff{3}Engineering Physics Department, University of Wisconsin--Madison, Madison, WI 53706, USA}

\begin{document}

\maketitle

\begin{abstract}
Quasi-periodic plasmoid formation at the tip of magnetic streamer structures is observed to occur in experiments on the Big Red Ball as well as in simulations of these experiments performed with the extended-MHD code, NIMROD. 
This plasmoid formation is found to occur on a characteristic timescale dependent on pressure gradients and magnetic curvature in both experiment and simulation.
Single mode, or laminar, plasmoids exist when the pressure gradient is modest, but give way to turbulent plasmoid ejection when the system drive is higher, producing plasmoids of many sizes. However, a critical pressure gradient is also observed, below which plasmoids are never formed. A simple heuristic model of this plasmoid formation process is presented and suggested to be a consequence of a dynamic loss of equilibrium in the high-$\beta$ region of the helmet streamer. This model is capable of explaining the periodicity of plasmoids observed in the experiment and simulations and produces plasmoid periods of 90 minutes when applied to 2D models of solar streamers with a height of $3R_\odot$. This is consistent with the location and frequency at which periodic plasma blobs have been observed to form by LASCO and SECCHI instruments.  
\end{abstract}

\section{Introduction}
Over the past few decades, \textit{in-situ} measurements of the solar wind have produced an enormous amount of data that can be used to characterize properties of the solar wind and to discover its source regions on the Sun. One of the earliest characterizations was the observation of ``fast'' and ``slow'' streams of wind as measured by Mariner II~\citep{Neugebauer1962}. However, later on it was discovered that the slow wind was better characterized by the charge state ratios of oxygen --- indicating a much higher electron temperature in the source region~\citep{Neugebauer2016, Fu2017, Cranmer2017}, consistent with the temperatures and charge states in well confined coronal loops. This led to the understanding that the slow solar wind likely originates from transport between closed flux and open flux in the equatorial streamer belt and that it can only occur via magnetic reconnection. This formation process is drastically more complex than the fast wind acceleration in coronal holes, which agrees with the original Parker solution \citep{Parker1958a} and produces a relatively quiescent, supersonic flow with photospheric abundances. Invoking magnetic reconnection in the formation mechanism of the slow wind inherently leads to a dynamic process capable of explaining its high degree of variability. However, spontaneous magnetic reconnection is difficult to achieve in high Lundquist number plasmas and so a mechanism with enough free energy for facilitating or driving the reconnection must be included in the theory. To this end, a number of theories have been postulated including ``interchange reconnection''~\citep{Crooker2000,Fisk1998,Fisk2003}, the S-Web theory~\citep{Antiochos2011,Higginson2018,Antiochos2007}, and streamer top reconnection~\citep{Einaudi2001,Einaudi1999,Lapenta2005,Endeve2004,Wu2000}. 

Specifically with regards to streamer top reconnection, a number of computational studies have attempted to recreate these periodic density structures (PDSs). This process can be driven by instabilities in the coronal loop or by converging flows at the streamer cusp~\citep{Einaudi2001,Einaudi1999,Lapenta2005,Wu2000} and has also revealed that two-fluid effects can alter plasmoid characteristics emanating from helmet streamers \citep{Endeve2003, Endeve2004}. These multi-fluid simulations prescribe a fixed amount of coronal heating at the base of the helmet streamer, which results in a dynamic system with no equilibrium that oscillates periodically. However, the periodicity of the plasmoids in these simulations is longer than that observed, $\sim$15-20 hours. 

With the improvements to imaging diagnostics on many of the current satellite missions (SOHO, STEREO, Parker Solar Probe), as well as novel data analysis techniques, increasingly smaller and more dynamic features are constantly being revealed in the solar wind~\citep{DeForest2018, Bale2019}. One example of this pertains specifically to the slow solar wind and the observation of PDSs, also known as plasma blobs or plasmoids, that are released into the solar wind at the tips of helmet streamers~\citep{Wang1997,SheeleyJr.1997,Lavraud_2020}. Running difference calculations of white light images produced by SOHO's Large Angle and Spectrometric Coronograph (LASCO)~\citep{Brueckner1995} reveal bipolar signatures indicative of small scale structures propagating outwards into the solar wind~\citep{Wang1998}. Recently these PDSs have also been identified by the SECCHI instrument suite onboard STEREO~\citep{Viall2015}, in old Helios data~\citep{DiMatteo2019}, and during Parker Solar Probe's first orbit~\citep{Lavraud_2020}. They have also been observed to have magnetic signatures~\citep{Rouillard2011} and be the product of magnetic reconnection at the open-closed flux boundary of helmet streamers~\citep{Kepko2016, Sanchez-Diaz2019}. 

While the work of \cite{Viall2015} shows that blobs form at or below $2.5 R_\odot$ and have a typical period of 90 minutes with a range of 65-100 minutes, other work suggests that blobs can also form at larger radii ($2-6 R_\odot$) and have longer periods of a few hours~\citep{Wang2018,Wang1998}. The implied correlation between these observations is that when helmet streamer tips are closer to the Sun they release higher frequency PDSs, and lower frequency PDSs when they are further away. This hypothesis is consistent with the observations of a wide number of variable discrete frequencies that are observed in the slow solar wind at 1AU over the course of the solar cycle~\citep{Viall2008}. 

Aside from the presence of multiple coherent frequencies of observed PDSs, it is also well understood that the heliospheric current sheet (HCS) --- and the solar wind as a whole --- is extremely turbulent~\citep{Bavassano1997,Coleman1968,Bavassano1989,Bavassano1989a,Luttrell1987,Belcher1971,Marsch1990a,Marsch1990b,Bale2019} and evolves as a function of distance from the Sun~\citep{Bavassano1982}. Even though fully developed turbulence is typically observed to exist by $0.3 R_\odot$ in the slow wind near the HCS, it is often not enough to completely decorrelate the coherent PDS fluctuations generated in the corona as they are routinely observed to drive magnetospheric fluctuations at 1AU~\citep{Viall2008,Kepko2002,Stephenson2002,Kepko2003}. 

The conclusion from this plethora of observational insight is that any mesoscale model that wishes to accurately describe the acceleration of the solar wind near the magnetic equator must be able to produce these coherent fluctuations embedded in a turbulent background that evolves as it travels away from the source. It is also necessary that the frequencies be on the proper time scales and that the fluctuations are consistent with density and magnetic signatures indicative of plasmoids.

In this article we report on experimental observations of axisymmetric plasmoid ejection via helmet streamer tip reconnection in a laboratory Parker Spiral. We discuss a potential plasmoid formation mechanism and observations of plasmoid scaling properties by comparisons between experimental measurements and extended-MHD simulations performed with the two-fluid modeling capabilities of the NIMROD code~\citep{Sovinec2004, Sovinec2010}. These scalings indicate higher frequency plasmoids for more elongated streamers with thinner current sheets, as well as an evolution toward turbulence further downstream as plasmoids of many sizes interact. 
Lastly, a simple heuristic model is presented suggesting that the periodicity of plasmoids in the experiments and simulations and those produced in the solar wind is set by a dynamic transition from a state of quasi-equilibrium to one where no such equilibrium exists --- a process we will refer to as equilibrium loss.
It is important to note that the helmet streamers produced in the lab and those in the solar wind represent drastically different systems; the former is governed by Hall-MHD and exists on sub-ion scales, whereas the latter is much better described by ideal MHD and exists on scales much, much larger than any kinetic scale. This article does not make any claims about the relationship of the dominant transport mechanisms between the two systems, nor does it claim that the underlying physics of the subsequent reconnection events are the same. The simple association is made that both systems exhibit sonic outflows that result in periodic plasmoid ejection from the tip of the helmet streamers and that the loss of equilibrium responsible for this phenomenon can be driven by pressure gradients and magnetic curvature both in the experiments and in the solar wind.

\section{Experimental and Simulation Methodologies}
\begin{figure}
\centering
\includegraphics[width=\textwidth]{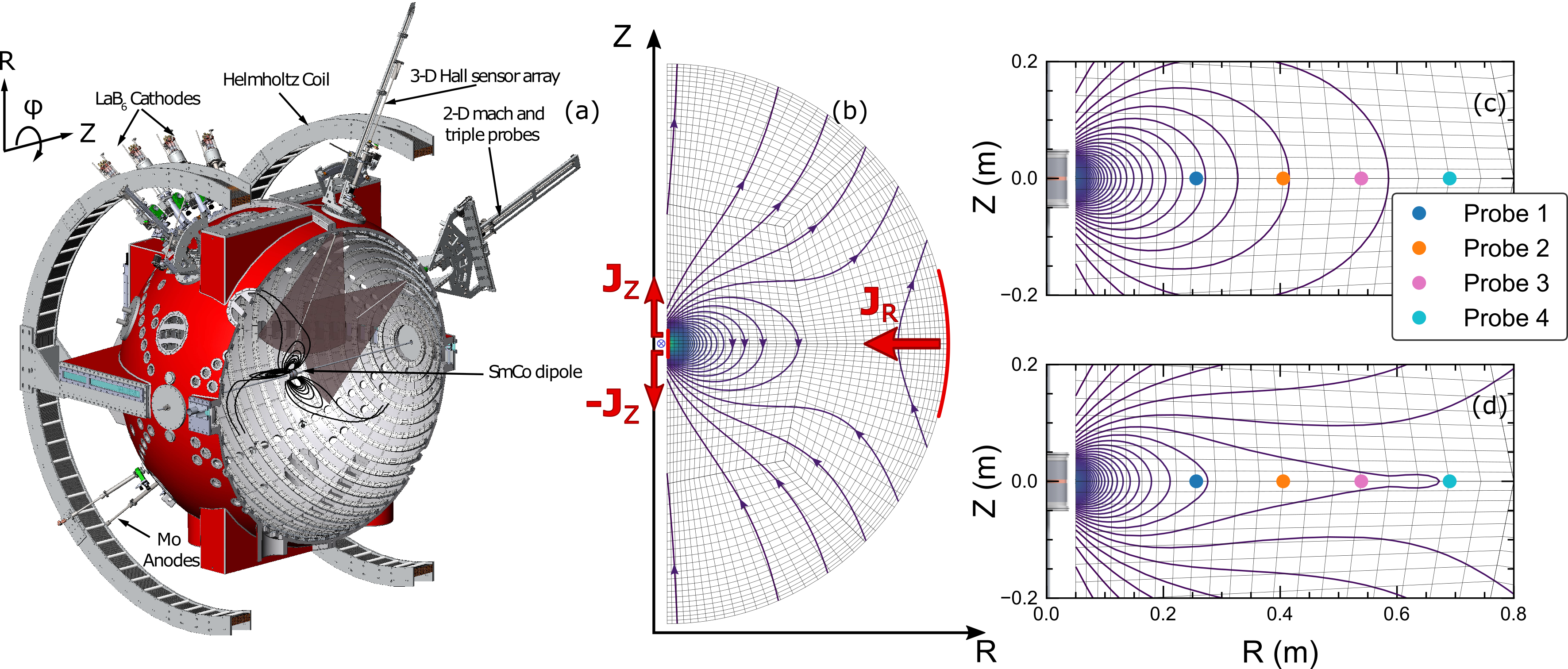}
\caption{Experimental (a) and computational (b, c, d) configurations. The Big Red Ball is shown in (a) along with the local cylindrical coordinate system, dipole magnet, and diagnostics used for mapping the magnetosphere. The 2D finite element mesh with current injection boundary conditions and initial vacuum field configuration are shown in (b). Panels (c) and (d) show the location of probes within the NIMROD simulation for outputting high time resolution field measurements. Both the initial magnetic field configuration (c) and time-averaged magnetic field configuration after the current injection has reached steady state (d) are shown to demonstrate the probe positions relative to where the plasmoid formation process occurs.}
\label{fig:setup}
\end{figure}

The Big Red Ball (BRB) at the Wisconsin Plasma Physics Laboratory is a versatile plasma confinement device well-suited to the study of high-$\beta$ and flow-dominated systems. The capabilities of the BRB are discussed in more detail in other works~\citep{Forest2015, Olson2016, Flanagan2020, Endrizzi2021} and the experimental setup of the Parker Spiral solar wind experiments and initial findings are detailed in \cite{Peterson2019}. A depiction of the experimental configuration of the BRB for these experiments as well as the computational model of the experiment are shown in \cref{fig:setup}. The summary of the experimental methodology for generating the Parker Spiral in the BRB is as follows: Thermally emissive Lanthanum-Hexaboride cathodes are used to generate a warm, dense, unmagnetized plasma atmosphere ($T_e \sim 7$ eV, $n_e \sim 4\times10^{17}$ m$^{-3}$). A permanent dipole magnet is placed inside this background plasma with two ring electrodes located near its north and south poles. Current is driven from molybdenum anodes in the plasma atmosphere into the dipole magnetosphere and collected on the polar electrodes. This current path is visualized in \cref{fig:setup}(b) in the context of the NIMROD simulation domain. These cross field currents generate a torque on the plasma which causes the magnetosphere to rotate, thereby twisting the magnetic field into a Parker Spiral. 

At the interface between the closed field lines of the magnetosphere and the open field lines of the Parker Spiral, periodic reconnection occurs, ejecting axisymmetric plasmoids, much like the density structures observed in the heliospheric current sheet which emanate from streamer top reconnection. A number of diagnostics were employed to map the 2D time dynamics of the Parker Spiral including two arrays of 3 axis Hall sensors and an array of triple probes and 2D Mach probes for measuring density, temperature, and flows in the plasma. One of the Hall sensor arrays was kept stationary and displaced azimuthally from the 2D scanning plane to provide phase reference measurements critical for the reconstruction of the plasmoid dynamics.

The computational domain and vacuum magnetic field for the accompanying NIMROD simulations are shown in \cref{fig:setup}. \Cref{fig:setup}(b) represents the hemispherical mesh in cylindrical coordinates, which extends down to $R=5$ cm rather than the experiment's support rod at $R=2$ cm. This is to allow for a small dipole magnet to be placed outside the computational domain and to facilitate the boundary condition manipulation to model current injection/extraction in a manner representative of the experiment. By prescribing $B_\varphi$ along the boundary as a function of time, we can set the normal component of $J$, or the current into and out of the vessel. In all the presented simulation work that follows, the current injection linearly ramps from zero up to a prescribed steady state value. The ramp duration in some of the initial simulations was 1 ms, but was shortened to 100 $\mu$s to reduce the required simulation time for some of the higher current injection cases that have smaller time steps. For all simulations discussed in this work, there are zero heat flux and zero particle flux conditions applied to the boundary as well as a no-slip boundary condition on the velocity. All simulations were performed with experimental parameters of $n_\mathrm{e} = 4\times 10^{17}$~m$^{-3}$, $T_\mathrm{e} = 7$~eV, $T_\mathrm{i} = 0.5$~eV, which give viscous and resistive diffusivities of $\nu = 50$~m$^2/$s and $\eta = 35$~m$^2/$s respectively. We also note that in the experiments and simulations the ion skin depth is set to $d_i = 70$ cm. In terms of resolution, the simulations were performed with 2400 bicubic poloidal elements which provides centimeter-scale resolution in the current sheet.

In order to better compare results from simulation to the experimental measurements, the NIMROD code was modified in order to take a list of R, Z coordinates for placing history nodes, or probes. For the simulations in this study, four probes were placed in the current sheet at $R=25, 40, 55, 70$ cm and $Z=0$ cm as shown in \cref{fig:setup}. The probe locations relative to the vacuum magnetic flux configuration and the flux distribution near the end of a simulation can be seen in \cref{fig:setup}(c) and \cref{fig:setup}(d) respectively. These probes provide outputs for the solution fields after every time step to allow for higher frequency phenomena to still be captured at a few select locations in the simulation. All NIMROD simulations presented in this work are axisymmetric.

\section{Observation of Streamer Top Reconnection and Plasmoid Formation in Experiment and Two-Fluid Simulations}
While the basic operating parameters, mean magnetic field, and plasma flows that develop in both the experiment and simulations are presented in \cite{Peterson2019}, this article sheds more light on the details of the fluctuation measurements as well as their scaling properties. Previous work demonstrated that, during the generation of the Parker Spiral, two fluid effects are critical as the system size is small compared to the ion skin depth, the electrons are the only magnetized species, and $T_e \gg T_i$. The consequence of this is a radially outward flow of electrons that advects the dipolar magnetic field into a Parker Spiral and simultaneously generates an inward electric field via the Hall effect. This Hall electric field drives accretion of ions into the magnetosphere where a density and pressure gradient begin to build until a loss of equilibrium occurs. In both experiment and simulation, this loss of equilibrium occurs in the current sheet associated with the Parker Spiral where $\beta > 10$ \citep{Peterson2019} and increases up to $\beta \sim 50$ further downstream.

\begin{figure}
\centering
\includegraphics[width=\textwidth]{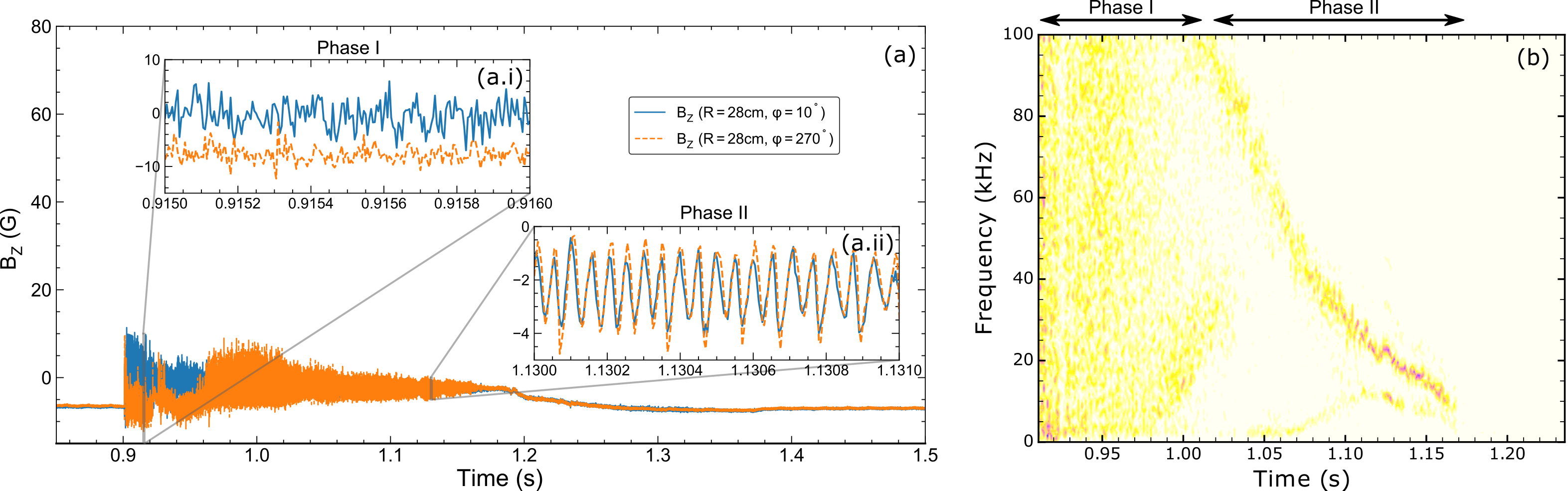}
\caption{Time histories and frequency content of magnetic signals measured in the Parker Spiral current sheet by 3-axis Hall sensors in the experiment. The difference between the turbulent, non-axisymmetric Phase I and axisymmetric, single mode Phase II can be seen in (a) where two probes at the same radius, but different azimuthal angles have very different mean field values as well as high frequency characteristics in (a.i.), but are nearly identical in Phase II (a.ii). A spectrogram of one of these time histories (b) shows broadband fluctuations in Phase I, followed by a coherent downward chirping spectrum of laminar plasmoid ejection.}
\label{fig:hall_traces}
\end{figure}

Measurements of the magnetic field as well as the plasma density show interesting fluctuations whose frequency depends on the amount of current injected. As shown in \cref{fig:hall_traces}(a), the vertical component of the magnetic field, $B_Z$ (the poloidal component perpendicular to the current sheet) is non-axisymmetric and highly uncorrelated in the initial, high current phase of the experiment (Phase I). However, as shown in \cref{fig:hall_traces}(a.ii), these magnetic fluctuations become coherent as the current injection falls with the characteristic timescale set by the RC circuit of the current injection system ($\sim 100$ms). In addition, the $B_Z$ fluctuations are bi-polar and thus suggestive of closed loops of magnetic flux disconnected from the inner magnetosphere of the rotating plasma. A spectrogram typical of these magnetic fluctuations in the current sheet can be seen in \cref{fig:hall_traces}(b) which shows the turbulent nature of the high current Phase I and coherent mode Phase II with a high degree of correlation between the plasmoid frequency and current injection of the system. This scaling relationship will be discussed in subsequent sections.

\begin{figure}
\centering
\includegraphics[width=.7\textwidth]{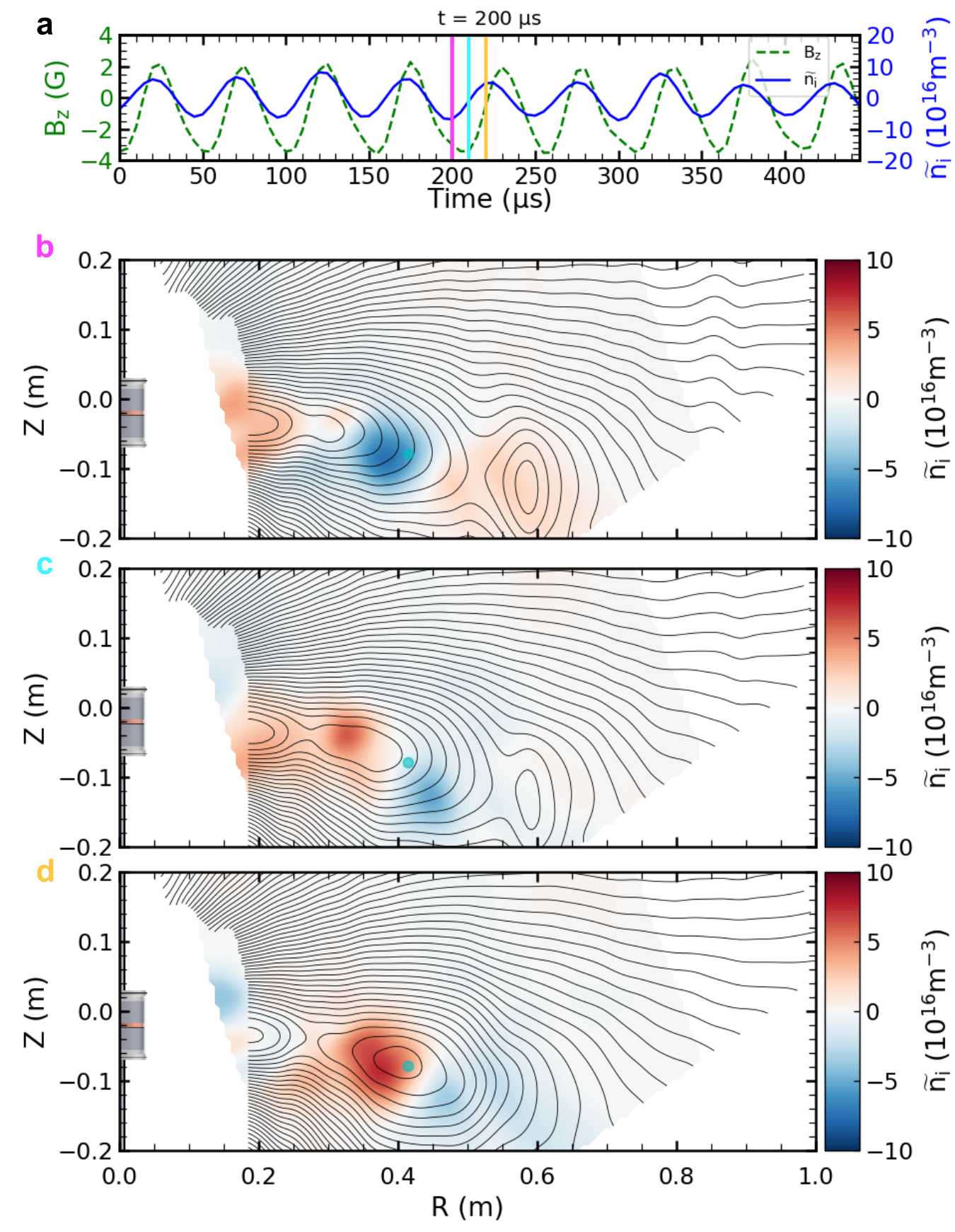}
\caption{Plasmoids are ejected from the helmet streamer cusp at a frequency of 20 kHz when the current injection is 150 A in the experiment. Panel (a) shows the magnetic field ($B_Z$) and density fluctuation ($\tilde{n}_i$) signals as measured by the probe in the current sheet at $R=42$ cm and denoted by the teal dot in panels (b), (c), and (d). Panels (b), (c), and (d) show the flux map and density perturbation map at three successive 10 $\mu$s time steps corresponding to the magenta, cyan, and yellow vertical lines in (a). These panels show how the magnetic field at the streamer cusp expands outwards with higher density plasma until reconnection occurs, releasing a plasmoid into the current sheet.}
\label{fig:exp_plasmoids}
\end{figure}
Performing phase correlation measurements between the stationary Hall array and the scanning array over many discharges allows us to reconstruct the plasmoid dynamics --- both magnetic field and density --- through conditional averaging. This reconstruction method first identifies 500 $\mu$s time windows for every shot that exhibit the highest degree of shot-to-shot similarity as measured by a single stationary hall probe. A center frequency of 20 kHz was used for this process as it showed the best correlation shot-to-shot between 1.1 and 1.15 seconds as shown in \cref{fig:hall_traces}(b). Once the best time window from each shot is selected, a phase shift is calculated for each shot using the stationary reference probe which allows us to align in time the many different shots at different locations and build up a poloidal map of the 2D plasma dynamics. This process reveals periodic reconnection occurring near the top of the streamer structures resulting in plasmoids being ejected into the Parker spiral current sheet as shown in \cref{fig:exp_plasmoids}. \Cref{fig:exp_plasmoids}(a) shows the $B_Z$ measurement as well as ion density fluctuations $\tilde{n}_i$ as measured by the probe located at ($R=42$cm, $Z=-8$cm) and shown as a cyan dot in \cref{fig:exp_plasmoids}(b-d). We believe the up-down asymmetry of the current sheet, or ``droop'', is likely due to preferential current draw to an anode in the southern hemisphere that was larger than the others, which is not captured in simulations since the current injection is uniform over the range of $\pm 15^\circ$ latitude. \Cref{fig:exp_plasmoids}(b-d) show subsequent time steps of 10$\mu$s through one half period of the reconnection process, which exhibits a periodic build up of plasma density (and pressure) inside the streamer leading to field line stretching, reconnection, and plasmoid ejection. 

\begin{figure}
\centering
\includegraphics[width=.8\textwidth]{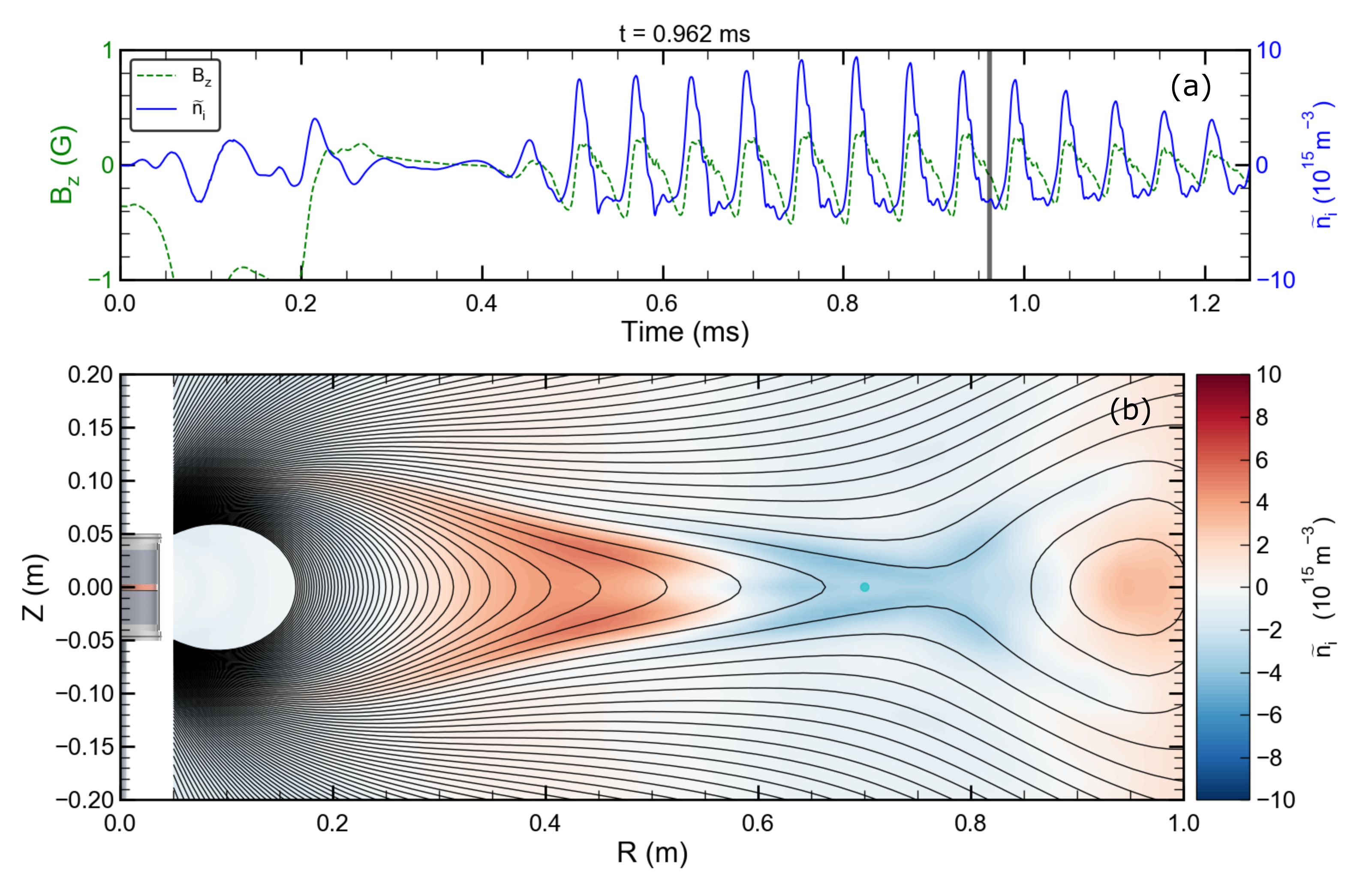}
\caption{Laminar plasmoid formation in an axisymmetric NIMROD simulation with 400A of injected current. The time history of $B_Z$ and ion density fluctuations as measured by a probe at $R=70$cm (top panel) shows the periodic ejection of high density plasmoids that are roughly similar size over time.}
\label{fig:nim_plasmoids1}
\end{figure}

\begin{figure}
\centering
\includegraphics[width=.8\textwidth]{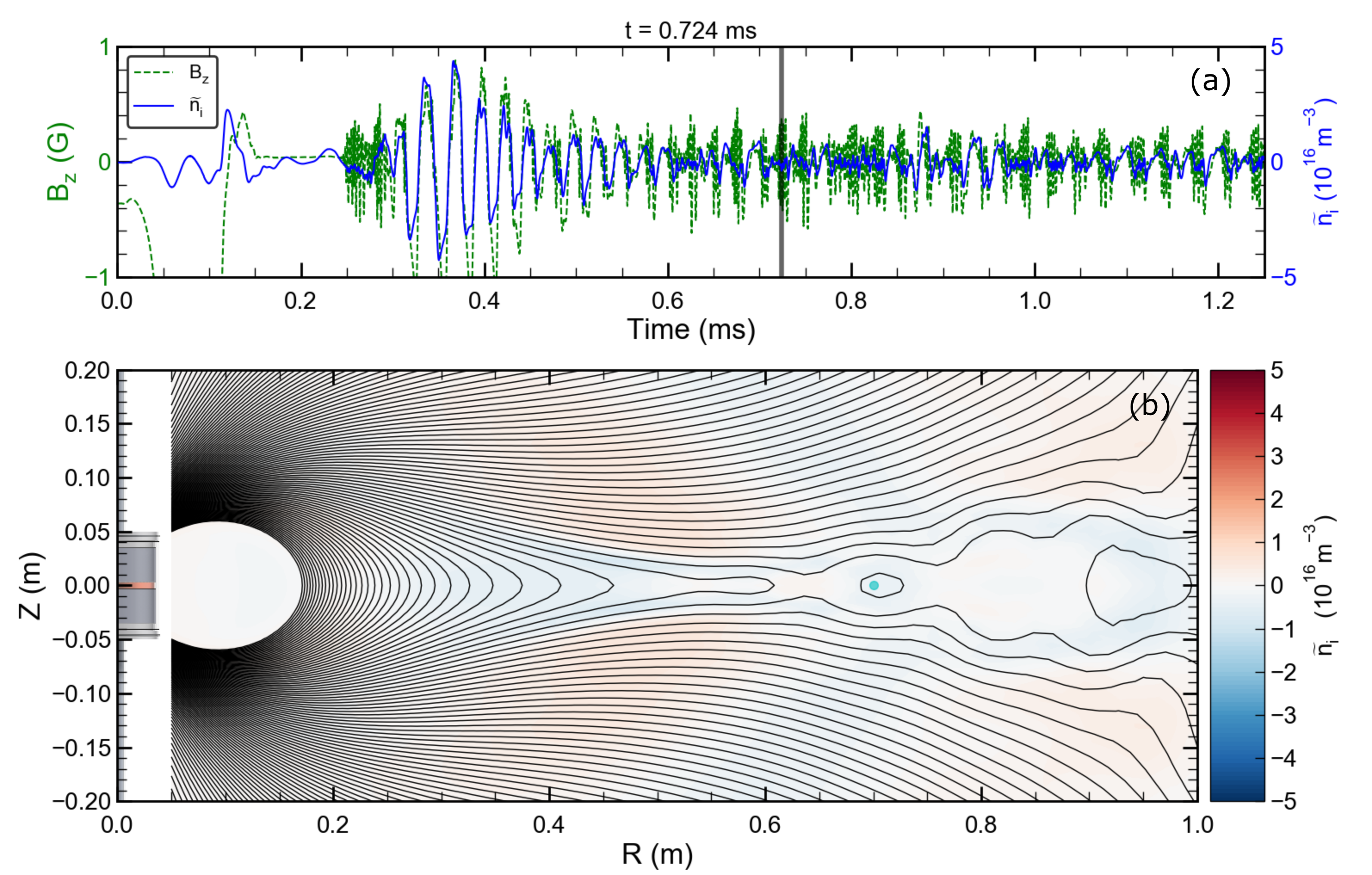}
\caption[Things]{Turbulent plasmoid formation in an axisymmetric NIMROD simulation with 1000A of injected current. The time history of $B_Z$ and ion density fluctuations as measured by a probe at $R=70$cm (top panel) shows the high variability of plasmoids in both frequency and size.}
\label{fig:nim_plasmoids2}
\end{figure}

In addition to plasmoids observed experimentally, they also manifest in extended MHD simulations when the Hall term and electron pressure gradient term are used in Ohm's law. 2D cross sections of the magnetic flux and density fluctuations from the NIMROD simulations as well as the time resolved signals in the current sheet as would be measured by probes in the experiment or satellites in space are shown in \cref{fig:nim_plasmoids1} and \cref{fig:nim_plasmoids2}. The time histories of both experimental and laminar plasmoids can be seen in movies published in previous work \citep{Peterson2019}, whereas the evolution of the turbulent plasmoids in \cref{fig:nim_plasmoids2} can be viewed in the movie \textit{turbulent\_plasmoids.mp4}. \Cref{fig:nim_plasmoids1} represents a moderate current injection level of 400 A in the simulation which is slightly above the threshold necessary to observe plasmoids. As a result, we observe relatively large single plasmoids emitted with a very regular frequency. We refer to these plasmoids as laminar since the magnetic flux evolves very smoothly with regular ejection of similarly sized plasmoids. 
On the other hand, \cref{fig:nim_plasmoids2} represents a high current drive case of 1000 A where we see plasmoids of many sizes interacting in a thinner current sheet resulting in a more turbulent medium downstream. 

The observation of plasmoids in the experiment as well as in the NIMROD simulations begs the question of whether this occurrence is coincidence or if the same physical processes are driving plasmoid formation in both cases. Importantly, theoretical and computational studies~\citep{Bhat2018}, as well as experiments~\citep{hare_2017} have demonstrated that plasmoid formation can occur at values of the Lundquist number much below  the usual $10^4$ required in resistive MHD when ion-scale kinetic effects are not negligible, as is the case here. We now turn our attention to a discussion of the properties and scaling relationships for these observed plasmoids.

\section{Discussion of Laminar and Turbulent Plasmoid Formation, Properties, and Scalings}

\begin{figure}
\centering
\includegraphics[width=\textwidth]{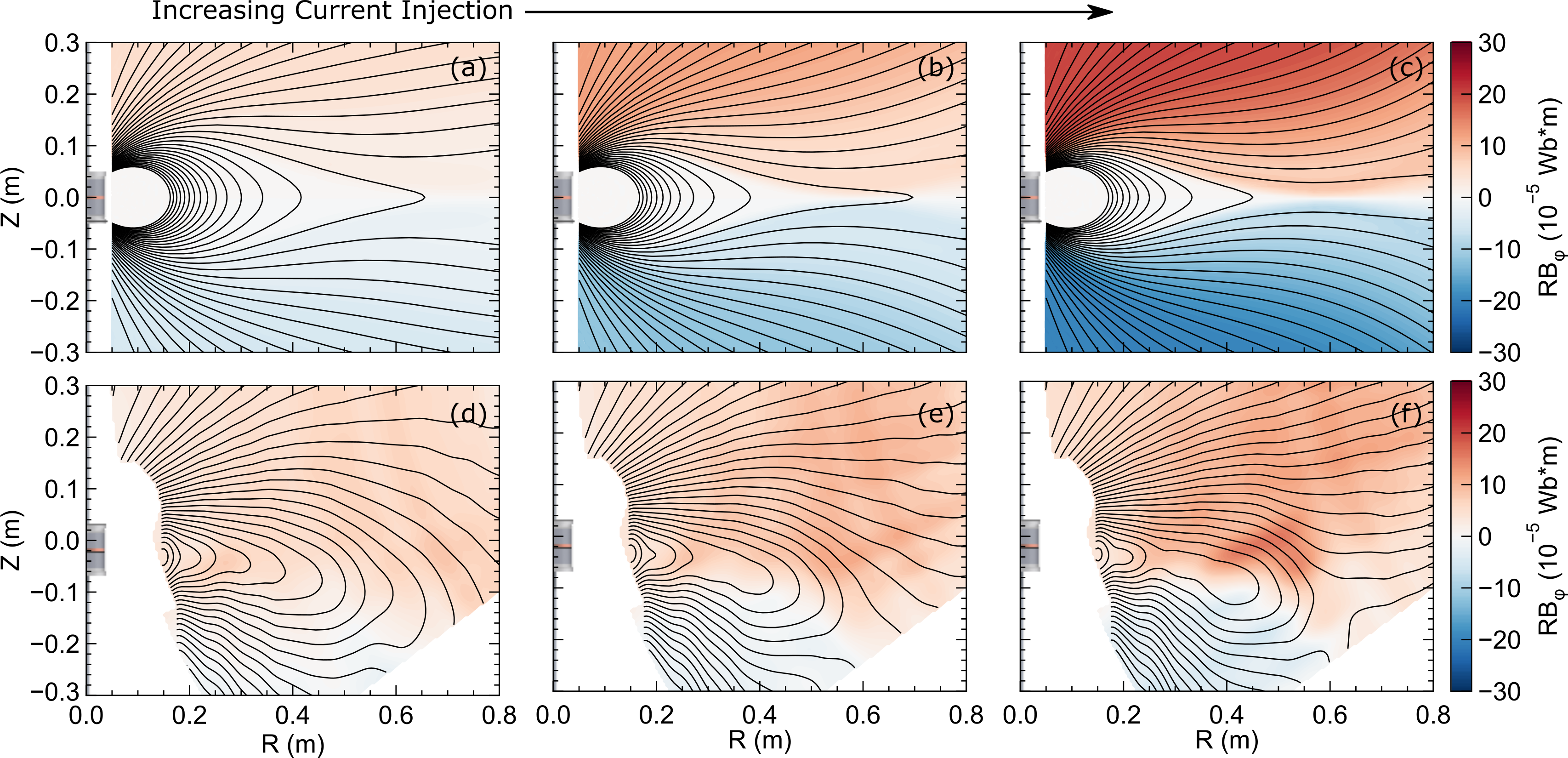}
\caption{Mean-field evolution of the magnetic field as a function of current injection is shown for axisymmetric NIMROD simulations (a-c) as well as the experiment (d-f). The top row depicts both the time-averaged poloidal and toroidal magnetic field at three different current injection values: 250 A, 600 A, and 1000 A, from left to right, respectively. In this progression, it is clear that the current sheet becomes thinner and the toroidal magnetic field increases as the poloidally injected current is increased. The same is true in the experiment as shown in the bottom row, where the current injection values are 150 A, 250 A and 350 A, respectively. For currents larger than 350 A in the experiment, the magnetic field is highly non-axisymmetric; as a result, axisymmetric flux surface reconstruction is not possible.}
\label{fig:eq_current_sheets}
\end{figure}

\begin{figure}
\centering
\includegraphics[width=\textwidth]{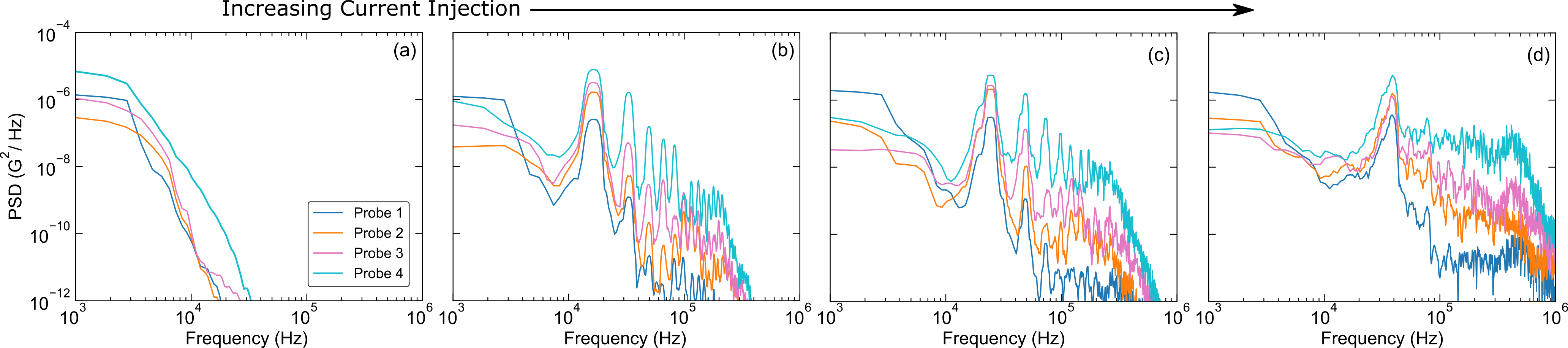}
\caption{$B_Z$ fluctuation power spectra from four probes in four different two-fluid NIMROD simulations with current values of 200A, 400A, 600A, and 1000A shown in panels (a), (b), (c), and (d), respectively. Probes 1-4 are located at increasing radii in the current sheet as according to \cref{fig:setup}. Increasing the injected current increases both the fundamental plasmoid frequency as well as the amplitudes of higher frequency components. Fluctuations are increasingly more broadband at larger radial distances as well.}
\label{fig:nimrod_probe_spectra}
\end{figure}

\begin{figure}
\centering
\includegraphics[width=\textwidth]{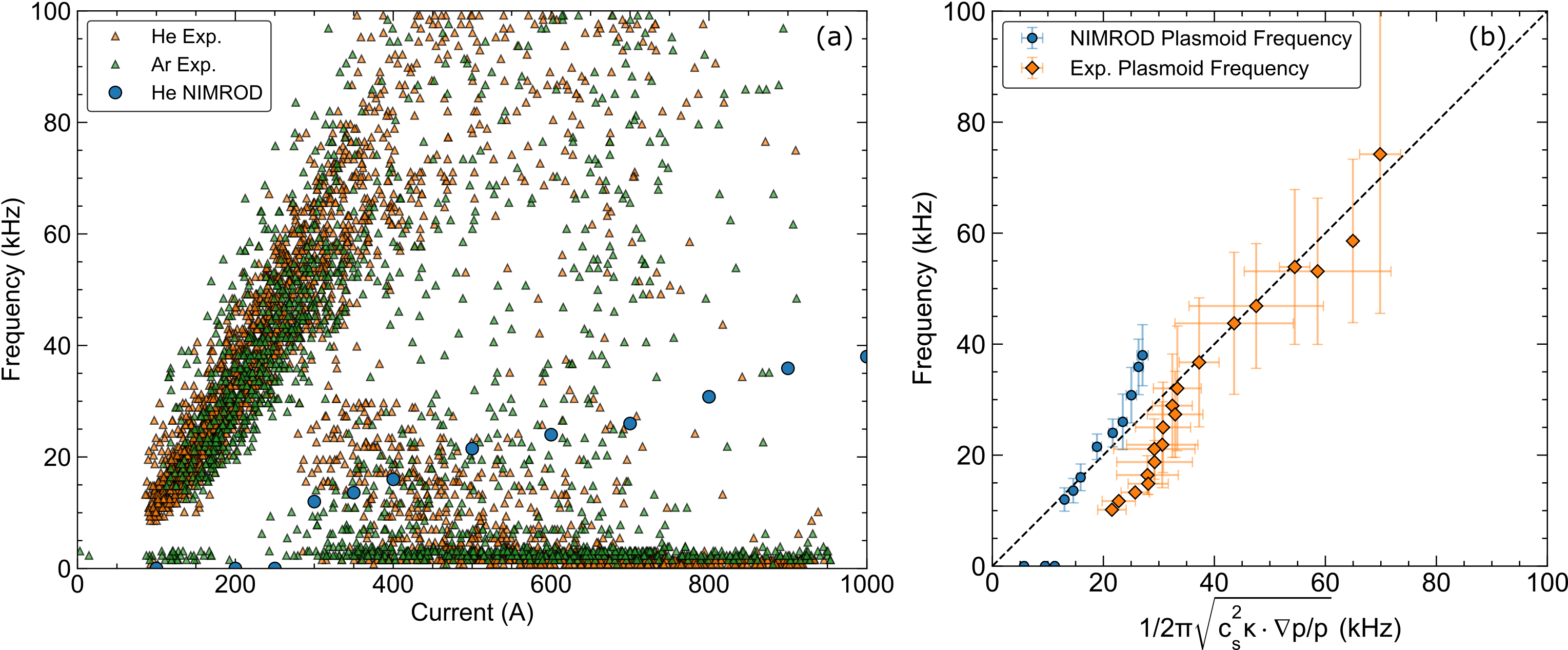}
\caption{$B_z$ fluctuation peak frequencies in the current sheet at $R=30$ cm for Helium and Argon discharges as well as the frequencies of plasmoids present in the NIMROD simulations (a). Plasmoids in both experiment and simulation scale linearly with the injected current and exhibit some stabilizing effect such that no plasmoids exists below roughly 10 kHz in either the experiment or simulations. When the current injection value is translated into a characteristic timescale dependent on the magnetic curvature and pressure gradient at that current value, the plasmoid frequencies in both experiment and simulation are found to scale similarly (b).}
\label{fig:freqs_vs_current}
\end{figure}

We begin by discussing the characteristics of these plasmoids, particularly with respect to the amount of current injected into the experiment (or simulation). Shown in \cref{fig:eq_current_sheets} are the time-averaged magnetic field configurations in the simulations (a-c) and experiment (d-f) as a function of increasing current injection. As more flux is advected outwards with the electron flow, it results in a current sheet thinning effect as shown in \cref{fig:eq_current_sheets}, where the thinner current sheets display larger magnetic curvature near the streamer cusp. As shown in both \cref{fig:nimrod_probe_spectra} and \cref{fig:freqs_vs_current}, these elongated streamers associated with higher current injection values result in higher frequency plasmoid ejection. However, plasmoids are not present for every current injection value in the simulation and experiment. This is shown in \cref{fig:nimrod_probe_spectra} which shows $B_Z$ power spectral densities from four different simulations with increasing amounts of current injection: 200A, 400A, 600A, and 1000A for panels (a), (b), (c), and (d), respectively. We can see that no plasmoids are present in the 200A simulation, likely because the accretion caused by the Hall effect is not strong enough to build up pressure above the critical gradient necessary for loss of equilibrium.

Another characteristic that trends with increased system drive (or current injection) is a decrease in coherence of the fluctuations and increase in turbulence. In \cref{fig:nimrod_probe_spectra}(b) we can see a well defined fundamental frequency at 15 kHz as well as multiple resolved harmonics. However, as the current is increased in panels (c) and (d), the spectra become more broadband and the fundamental mode increases in frequency, which is consistent with the experimental observations reported in \cite{Peterson2019}.

One more important observation from \cref{fig:nimrod_probe_spectra} is that field lines at inner radii (closer to probe 1) are more coherent --- that is, the dominant mode is much stronger relative to the other higher frequencies. The physical interpretation of this phenomenon is that the pressure inside the magnetosphere drives a periodic loss of equilibrium that manifests as magnetospheric oscillations that drive larger fluctuations further out in the current sheet where the field is weaker, ultimately resulting in a turbulent current sheet that still has a quasi-periodic nature.

A comparison of the plasmoid frequencies in simulation to those observed in the experiment is shown in \cref{fig:freqs_vs_current}(a) as a function of current (orange and green triangles). In \cref{fig:freqs_vs_current}(a) each experimental data point represents the peak value in the $B_Z$ frequency spectrum from a single Hall probe in 10 ms windows between 0.9 and 1.4 seconds. This generates 50 data points per shot and is plotted for roughly 100 shots. The simulation results (blue circles) show the dominant frequency for the duration of the simulation once the current injection has reached steady state. Each data point for the NIMROD simulations therefore represents a single simulation at a single current injection value. It is important to note the strong linear scaling at modest current injection up to $\sim 400$ A, as well as the abrupt disappearance of plasmoids below 100 A and $\sim 10$ kHz. The high density of data points at very low frequency for currents above $\sim 300$ A, are not particularly germane to this discussion and just indicate that the frequency range with the highest power spectral density was found to be at low frequencies during the non-axisymmetric phase of the experiment and can be seen as well in the Phase II portion of the spectrogram in \cref{fig:hall_traces}(b). Also plotted is the fundamental plasmoid frequency from the Hall-MHD NIMROD simulations as a function of current. Both experimental and simulation plasmoid frequencies scale linearly with the current, but with different slopes.  

Since the plasmoid frequencies scale more strongly with current injection in the experiment than in simulation, the drive mechanism is likely correlated with some quantity other than the current, but likely influenced by it. One possible explanation is that these plasmoids are pressure driven and that the current drive in the experiment produces larger densities in the magnetosphere as a result of ionization, which is not modeled in the simulations. Therefore, calculating a pressure-curvature driven time scale for the loss of equilibrium in experiment and simulations may provide a unifying scaling, as evidenced by \cref{fig:freqs_vs_current}(b). Each data point for the experiment in \cref{fig:freqs_vs_current}(b) is derived from a 2D flux surface reconstruction and maps of pressure, and temperature averaged over 10 ms windows. Where the flux surface reconstruction is possible (roughly from $t=1.0$ s to $t=1.2$ s), we obtain an estimate of the characteristic frequency $f = \frac{1}{2\pi}\sqrt{c_s^2\vb*{\kappa}\cdot\grad{p}/p}$ calculated near the helmet streamer tip at $R \sim 30$ cm. This results in $\sim 20$ data points whose error in the x-direction is calculated from the error in the magnetic field, temperature, and pressure measurements and whose error in the y-direction represents the full width at half maximum of the peak in the $B_z$ power spectral density. The theoretical model and justification for this characteristic time scale are explained in the following section.

\section{Heuristic Model of Plasmoid Evolution and Extrapolation to Solar Streamers}
\label{sec:model}

\begin{figure}
    \centering
    \includegraphics[width=.75\textwidth]{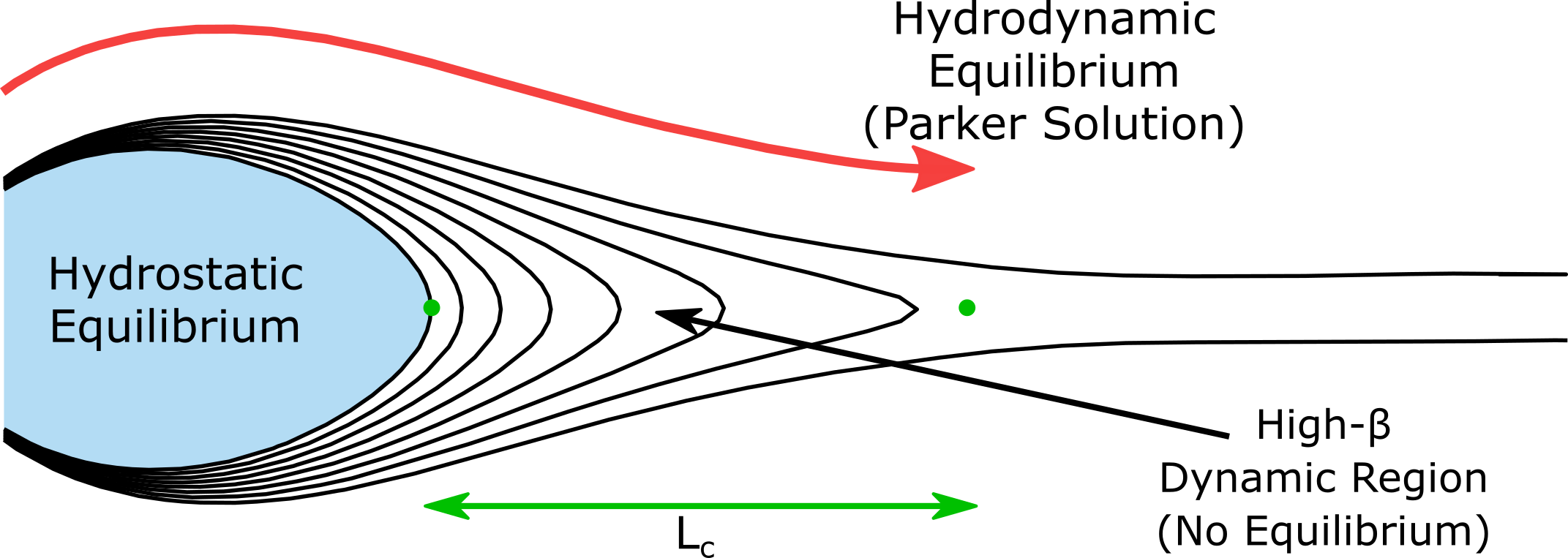}
    \caption{A cartoon drawing of helmet streamer topology with indications of the zones where equilibrium exists; namely, deep in the streamer where there is a hydrostatic solution, and outside the current sheet where Parker's original hydrodynamic solution is valid. The interface between the two exhibits no equilibrium due to the particle and heat sources present in the corona. This loss of equilibrium is caused by the extremely high values of $\beta$ in this region and result in quasi-periodic plasma blobs which are driven by strong pressure gradients. The length scale $L_c$ shown above is the critical length scale of the current sheet that develops from field line stretching before reconnection occurs.}
    \label{fig:streamer_cartoon}
\end{figure}

A relatively simple heuristic model can be constructed for the plasma expansion in the high-$\beta$ transition region between the hydrostatic equilibrium of the closed flux corona and the hydrodynamic equilibrium of open field lines outside the heliospheric current sheet (HCS). The geometry of such a system is shown in \cref{fig:streamer_cartoon}. Writing the momentum equation in terms of the magnetic curvature vector, $\vb*{\kappa}$, we obtain the dynamic equation for the plasma expansion:
\begin{equation}
    \rho\dv{\vb{v_\perp}}{t} = -\grad_\perp{p} + \frac{B^2}{\mu_0}\vb*{\kappa} - \grad_{\perp}\frac{B^2}{2\mu_0}.
    \label{eqn:expansion}
\end{equation}

Using the equation of state $p = \rho c_s^2$, rewriting the right hand side of \cref{eqn:expansion} in terms of the plasma $\beta = 2\mu_0 p / B^2$, and dotting both sides with the curvature vector, $\vb*{\kappa}$, gives:
\begin{equation}
    \vb*{\kappa}\cdot\dv{\vb{v_\perp}}{t} = -\frac{c_s^2}{p}\vb*{\kappa}\cdot\grad_\perp \left[p\left(1 + 1/\beta\right)\right] + \frac{2c_s^2}{\beta}\kappa^2.
    \label{eqn:kappa_dot_momentum}
\end{equation}

The left-hand side of this equation can be taken to define a characteristic timescale for the driven loss of equilibrium:  $\vb*{\kappa}\cdot\dv{\vb{v_\perp}}{t} \sim -\gamma_{dr}^2$. This can be understood as the time required to form a plasmoid of radius $R_c = \kappa^{-1}$ due to the acceleration provided by the net force on the right-hand side of \cref{eqn:kappa_dot_momentum}. As discussed before, the values of $\beta$ in the current sheet are very large for both the experiment and simulations and range from $10 < \beta < 50$ throughout the region of interest, denoted as the high-$\beta$ dynamic region in \cref{fig:streamer_cartoon}. Therefore, ignoring terms of order $\beta^{-1}$ results in 

\begin{equation}
    \gamma_{dr}^2 \sim \frac{c_s^2}{p}\vb*{\kappa}\cdot\grad{p}.
    \label{eqn:kdgp_gamma}
\end{equation}

This characteristic time scale can be understood as the time for a sound wave to traverse a distance equivalent to the geometric mean of the pressure gradient scale length and radius of curvature, or equivalently the free-fall time of a plasma parcel under the action of a pressure gradient and adverse magnetic field curvature. 

The next step in the derivation of the plasmoid frequency is to show that it is essentially this drive frequency which sets the frequency of reconnection in the current sheet. It will be shown that in this situation the reconnection rate is relatively insensitive to the details of the tearing mode growth rate. We can assume that the current sheet is lengthening at the rate given by the drive timescale $\gamma_{dr}$ because the plasma is frozen to the magnetic field. We can also assume that the lengthening of the current sheet is exponential in time: a result of a transition from quasi-equilibrium to a dynamic system. Incompressibility then requires that the forming sheet is likewise thinning exponentially, such that its thickness $a(t)$ can be described by 
\begin{equation}
    a(t) = a_0e^{-\gamma_{dr}t},
    \label{eqn:cs_thinning}
\end{equation}
where $a_0$ is the thickness at the beginning of the expansion.

As the aspect ratio $L/a$ of this forming current sheet increases, it becomes unstable to a progressively broader spectrum of tearing modes. As argued by~\citet{Uzdensky2016}, one of those modes --- the one whose growth rate, $\gamma_{tear}(t)$, first satisfies $\gamma_{tear}(t_{crit})\sim \gamma_{dr}$ --- will eventually grow to become as wide as the forming sheet, thereby disrupting it and leading to plasmoid ejection. This happens at a so-called critical time, $t_{crit}$, whereupon the sheet thickness is
\begin{equation}
    a_{crit} \equiv a(t_{crit})= a_0e^{-\gamma_{dr}t_{crit}}.
    \label{eqn:cs_crit}
\end{equation}
This can be inverted to yield
\begin{equation}
    t_{crit} = \gamma_{dr}^{-1} \ln{\left(\frac{a_0}{a_{crit}}\right)}.
    \label{eqn:t_crit}
\end{equation}
As we can see, the critical time for reconnection to occur is essentially the same as the drive timescale, $\gamma_{dr}^{-1}$, since the ratio of the initial to the critical current sheet thickness represents only a logarithmic correction. That is, while reconnection is essential to the formation and ejection of plasmoids, the physics of the reconnection onset are such that details of the tearing instability that underlies it (such as the functional form of $\gamma_{tear}(t)$) are not essential to the prediction of the timescale associated with the plasmoid ejection.

Casting the drive timescale for $\gamma_{dr}$ into a characteristic frequency in Hz we obtain
\begin{equation}
    f = \frac{1}{2\pi}\sqrt{c_s^2\vb*{\kappa}\cdot\grad{p}/p}.
    \label{eqn:kdgp_f}
\end{equation}

This time scale is associated with a high-$\beta$ pressure-curvature driven loss of equilibrium. The characteristics of these oscillations --- namely that they are electromagnetic, axisymmetric perturbations localized to the region of bad magnetic curvature with a frequency dependent on the pressure gradient --- support the notion that these plasmoids are driven by a mechanism similar in nature to flute modes, but likely represent a loss of equilibrium due to particle and heat transport into the streamer rather than a linear instability. It is the frequency in \cref{eqn:kdgp_f} that empirically provides the unifying scaling between the experimental results and the extended-MHD simulations presented in \cref{fig:freqs_vs_current}. Computing this frequency along field lines from the density, temperature and magnetic flux measured by diagnostics in the experiment as well as the simulation shows a local maximum located at the outboard midplane where the curvature is largest. Plotting the measured plasmoid frequencies against this calculated pressure-curvature frequency gives the results in \cref{fig:freqs_vs_current}(b) which shows much better agreement between the experiment and simulations and is consistent with the idea that these plasmoids are pressure driven.

If we normalize both the pressure gradient scale length and the magnetic curvature by the critical current sheet length, $L_c$, we obtain two dimensionless quantities: $\ell_b^{-1} = L_c\kappa$ and $\ell_p^{-1} = L_c\grad{p}/p$. We can assume, in the laminar plasmoid case, that this critical length, $L_c$, is simply the plasmoid length just as reconnection occurs. Substituting these parameters into \cref{eqn:kdgp_f} provides us with the dimensional scaling below:

\begin{equation}
    f = \frac{1}{2\pi}\frac{c_s}{L_c\sqrt{\ell_b\ell_p}}.
    \label{eqn:kdgp_f_dim}
\end{equation}

Given similar normalized length scales, $\ell_b$ and $\ell_p$, the scaling between experimental and solar wind frequencies is simply the ratio of critical length scales, or plasmoid size, and sound speeds; i.e.: 
\begin{equation}
    f_{sw} \sim \frac{L_{c,exp}}{L_{c,sw}}\frac{c_{s,sw}}{c_{s,exp}}f_{exp}.
\end{equation}

As mentioned before, we will use the plasmoid length as the proxy for $L_c$ as it is reasonable to assume for single plasmoids that the associated plasmoid is roughly the size of the current sheet
just before it reconnects. For plasmoids in the simulations and experiment, we will take this length scale to be $L_{c,exp} = 0.25$ m and in the solar wind we will take this scale to be $L_{c,sw} = 1 R_\odot = 7\times10^8$ m which is consistent with observations of plasmoids appearing around $2 - 4 R_\odot$ and having a length of $1 R_\odot$ and width of $0.1 R_\odot$\citep{Sheeley2009}. Combining these plasmoid length scales with the sound speed typical of the experiment ($c_{s,exp} \sim 13$ km/s) and solar corona ($c_{s,sw} \sim 200$ km/s) gives us a simple scaling relationship between frequencies observed in the lab and in the solar wind as shown in \cref{eqn:fsw}
\begin{equation}
    f_{sw} \sim 5.5\times 10^{-9}f_{exp}.
    \label{eqn:fsw}
\end{equation}

Therefore, the $20 - 40$ kHz plasmoids observed in both the experiment and simulations correspond to a plasmoid frequency in the solar wind of $110 - 220~\mu$Hz, or periods of $75 - 150$ minutes --- in remarkable agreement with observations \citep{Viall2015}.

We note that this model also offers a natural explanation for the existence of laminar and turbulent plasmoid regimes. Increasing the drive (i.e., increasing the injected current in experiments and simulations) leads to a larger $\gamma_{dr}$. When the drive is strong enough such that $\gamma_{dr} \gtrsim v_A / L_c$, an ejected plasmoid has insufficient time to advect downstream a distance comparable to its length before the subsequent plasmoid is ejected. This leads to plasmoids of different sizes interacting downstream of the reconnection region and more stochastic behavior. For the high drive cases in the simulation, many of the plasmoids are small ($L_c \sim 5$ cm), and $v_A \sim 2$ km/s, which results in the condition $\gamma_{dr} \gtrsim 40$ kHz which is in good agreement with the onset of the turbulent plasmoids shown in \cref{fig:nim_plasmoids2} and \cref{fig:nimrod_probe_spectra}(d). 

Another mechanism that may be contributing to the turbulent dynamics and to the non-axisymmetric nature of Phase I in the experiment is the transition from single plasmoid formation to multiple simultaneous plasmoid formation. To support this hypothesis, the calculation in \cref{sec:appendix} computes the threshold drive frequency necessary to destabilize shorter wavelength tearing modes in the current sheet. If this threshold is reached we may conclude the subsequent stochastic dynamics are those of a plasmoid chain~\citep{uzdensky_2010,loureiro_2012} and are responsible for the turbulence in the high-drive cases. Remarkably, as outlined in \cref{sec:appendix}, this threshold frequency is computed to be $f=\gamma_{dr}/(2\pi) \gtrsim 60\,$kHz and is precisely in the range we would expect based on experimental evidence ($\sim 60-100$ kHz). 

These two mechanisms allow us to infer a hierarchy of stochasticity associated with the drive strength and therefore the frequency of plasmoid formation. As the drive strength increases, the system experiences a loss of equilibrium and laminar `single' plasmoid ejection. This phase is followed by plasmoids of different `generations' catching up with each other to generate a turbulent medium downstream, but still in the `single' plasmoid regime. This phase applies particularly to the NIMROD simulations which are constrained to be axisymmetric and are below the threshold for multiple plasmoids, yet still exhibit increased turbulence at higher drive. Finally, at the highest drives obtained in the experiment, the threshold to multiple simultaneous plasmoid formation is reached as discussed in \cref{sec:appendix}.

This analysis can be made more quantitative by formally taking into account the geometric effects of the magnetic field and pressure gradient for helmet streamer structures in the solar wind. This process likewise results in blob periodicities in the 1-2 hour range consistent with observations \citep{Viall2015}. This model is constructed by taking the streamer-like poloidal field geometry generated during NIMROD simulations and scaling the radius where reconnection occurs to coincide with $3 R_\odot$ as a characteristic location for PDS formation \citep{Viall2015, Wang1998, Wang2018}. Scaling the magnetic field strength using fits to solar wind data in the ecliptic plane according to \cite{Kohnlein1996}, produces a plausible helmet streamer geometry at the proper scale with realistic magnetic field strength. Using fits for plasma temperature and density in the ecliptic (likewise from \cite{Kohnlein1996}) and mapping them to the respective flux surfaces produces a mock helmet streamer with plausible temperature and density profiles. 

\begin{figure}
    \centering
    \includegraphics[width=.95\textwidth]{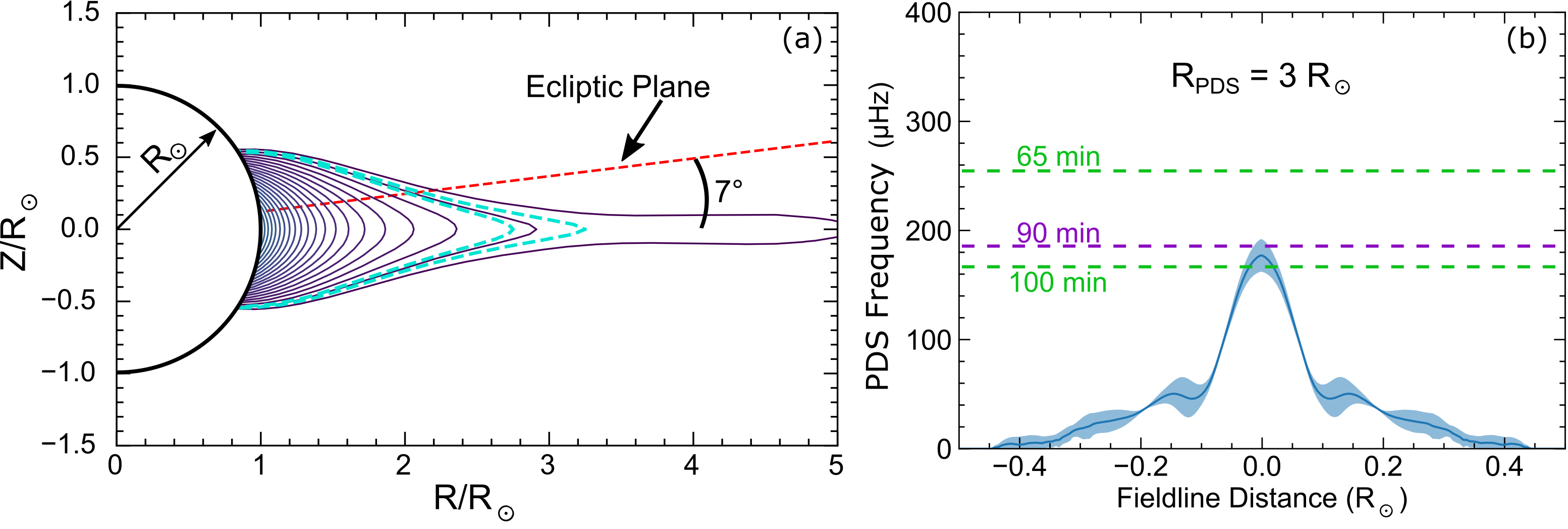}
    \caption{The work of \cite{Kohnlein1996} provides doubly logarithmic fits to Helios data for density, temperature, and magnetic field as a function of heliocentric distance in the ecliptic plane. Mapping these quantities onto the magnetic streamer structure shown in (a) allows us to compute $f = 1/2\pi\sqrt{c_s^2\vb*{\kappa}\cdot\grad{p}/p}$ along field lines just inside and just outside the reconnection radius in the same fashion as shown in \cref{fig:freqs_vs_current}(b). This results in plasmoid frequencies that are peaked at the streamer cusp as expected and produce periodicities in close agreement with \textit{in situ} observations for plausible solar wind parameters.}
    \label{fig:streamer}
\end{figure}
A diagram of this 2D magnetic geometry for this heuristic model is shown in \cref{fig:streamer}(a). This model enables us to calculate the same characteristic frequency used to unify the scaling between experiment and simulation (\cref{eqn:kdgp_f}) along the field lines in the vicinity of the reconnection site shown as the cyan dashed lines in \cref{fig:streamer}(a). The result of calculating this frequency along flux surfaces between the cyan dashed flux surfaces in \cref{fig:streamer}(a) is shown in \cref{fig:streamer}(b) and plotted as a function of field line distance away from the outboard midplane. In this context, a field line distance of 0 corresponds to the streamer top or the point of highest magnetic curvature where we might expect the highest growth rate of any pressure-curvature driven loss of equilibrium. We can see from this model that the same characteristic timescale used to unify experimental and simulation plasmoid frequencies results in a blob periodicity in the solar wind of $\sim90$ minutes if formed at $3 R_\odot$ (\cref{fig:streamer}b). If the PDS origin radius is increased or decreased, the PDS periodicity likewise increases or decreases respectively in accordance with observations \citep{Wang2018}.

\section{Conclusions}
To accompany the experimental measurements of streamer top reconnection and plasmoid formation in the Parker Spiral current sheet, a wide range of extended-MHD simulations were performed with the NIMROD code. Through measurements and comparisons between experiment and simulation, we showed that this high-$\beta$ loss of equilibrium is related to the pressure gradient and magnetic curvature in the streamer. Specifically, the frequency of expelled plasmoids scales with the pressure gradient and becomes more turbulent as the pressure gradient increases and as one moves further downstream in the current sheet. 

A heuristic model for this loss of equilibrium is presented and demonstrates that pressure-curvature driven outflows in the high-$\beta$ transition region of the streamer belt may be responsible for the streamer top reconnection that fuels a portion of the slow solar wind near the HCS. Although the parameters and scale lengths in the experiment are considerably different from those in the solar wind, the pressure driven loss of equilibrium allows both systems to expand outwards at their respective sound speed, advecting the magnetic flux with the ions in the case of the solar wind and with the electrons in the experiment and simulations. While the dynamics of magnetic reconnection are likewise vastly different between the two systems --- occurring on sub-ion scales in the experiment and macroscopic (MHD) scales in the solar wind --- it is likely that the plasmoid formation rate is governed more by the drive timescale, $\gamma_{dr}$, than by the specifics of magnetic reconnection. As a result, the streamer top may be reconnecting at a rate governed by the particle and heat sourcing on these outer field lines which results in loss of equilibrium rather than a linear instability. This allows for a unified theory to connect observations in drastically different regimes of plasma physics based on empirical evidence. While the underlying reconnection dynamics which set the critical length scale of the current sheet are certainly different, the resulting phenomenon was found to be remarkably similar between experiment and simulation and was also reminiscent of observations of the solar corona performed by the LASCO and SECCHI instrument suites as well as recently by Parker Solar Probe~\citep{Lavraud_2020}. 

\bigskip
The present work was supported by the NASA Earth and Space Sciences - Heliophysics Division Fellowship award no. NNX14AO16H. The BRB facility was constructed with support from the National Science Foundation and is now operated as a Department of Energy National User Facility under DOE fund DE-SC0018266. In addition, this work was supported by the NSF-DOE Partnership in Basic Plasma Science and Engineering award no. PHY-2010136.

\appendix
\section{Reconnection onset in a forming current sheet in the collisional Hall-MHD regime}
\label{sec:appendix}

In Section~\ref{sec:model} we alluded to the onset of the tearing mode in the forming current sheet driven by the equilibrium loss. In this Appendix, we present a quantitative, though simplified, derivation aimed at capturing what we think are the key features of this process in our experiments and simulations.

The plasma regime of relevance here can be described by the resistive Hall-MHD framework; namely, we take the ions to be cold, the reconnection dynamics to be happening at sub-ion-skin-depth scales, and the frozen-flux condition to be broken by resistivity. 
Under these constraints, the expressions for the growth rate of the tearing instability for small and large values of the tearing instability parameter $\Delta'$ can be obtained from~\citet{Attico_2000}\footnote{\citet{Attico_2000} consider the case where resistivity is negligible and the frozen flux constraint is instead broken by electron inertia. The resistive scalings that we use here are directly retrievable from theirs upon the substitution $d_e^2\rightarrow \eta/\gamma$, where $d_e$ is the electron skin depth.}. 
They are
\begin{equation}
\label{eq:low_delta}
    \gamma \tau_w = 0.47\, \Delta' (\tau_w \eta)^{1/2},
\end{equation}
in the low $\Delta'$ case; and
\begin{equation}
\label{eq:large_delta}
    \gamma \tau_w = 0.69\, (k a)^{3/4} a^{-1/2} (\tau_w \eta)^{1/4},
\end{equation}
in the large $\Delta'$ case. The normalizing timescale that appears in these expressions is sometimes called the whistler time, $\tau_w = a^2/(d_i v_A)$, with $v_A$ the Alfv\'en speed based on the upstream (reconnecting) magnetic field.

Given a spectrum of unstable wavenumbers, the fastest growing tearing mode is given by the intersection of these two scalings:
\begin{equation}
\label{eq:kmax}
    k_{max}a \approx 1.2\, (\tau_w \eta)^{1/7} a^{-2/7},
\end{equation}
where we have assumed that the upstream magnetic field is well represented by a $\tanh{x/a}$ profile, whose instability parameter is $\Delta' a \approx 2/(k a)$ for $ka\ll 1$. 
The corresponding growth rate is
\begin{equation}
    \gamma_{max}\tau_w \approx 0.8 (\tau_w \eta)^{5/14} a^{-5/7}.
\end{equation}
This mode is the fastest-growing mode if the current sheet is long enough that it fits inside the layer; i.e., if $k_{max} L\ge 1$.

From here, the calculation proceeds exactly as prescribed in~\citet{Uzdensky2016}. We assume, as in Section~\ref{sec:model}, that the length and the thickness of the forming current sheet expand, or contract, exponentially, with the drive rate $\gamma_{dr}$. 
Then we find that the $N=1$ mode transitions from the low to the large $\Delta'$ regime at the time
\begin{equation}
    t_{tr} = \frac{1}{2\gamma_{dr}} \ln \left(\frac{2\pi}{1.2}\frac{a_0}{L_0} S_H^{1/7} \right),
\end{equation}
where $S_H \equiv d_i v_A/\eta$ is the Hall Lundquist number, and $a_0$ and $L_0$ are the initial thickness and length of the current sheet.

On the other hand, one can compute the time $t_{cr}$ at which the growth rate of the $N=1$ mode matches the current sheet formation rate; i.e., solve $\gamma(t) = \gamma_{dr}$ for the $N=1$ mode. This yields
\begin{equation}
    t_{cr} \approx \frac{1}{4\gamma_{dr}} \ln \left(\frac{\pi}{0.47}\frac{a_0}{L_0}\frac{a_0^2 \gamma_{dr}}{d_i v_A}S_H^{1/2} \right).
\end{equation}

Finally, one can ask if the $N=1$ mode has the time to transition from the low to the large $\Delta'$ regime before reaching its critical time. This occurs when
\begin{equation}
    \gamma_{dr}> 4.1 \frac{a_0}{L_0} \frac{d_i v_A}{a_0^2} S_H^{-3/14}.
\end{equation}
In other words, if this condition is satisfied, one would expect the forming current sheet to be disrupted by multiple plasmoids (large tearing mode number, $N$); if it is not, then it is the $N=1$ tearing mode that disrupts the forming sheet.

Inserting the values measured or inferred from the experiments into this expression (namely, $a_0=0.05\,$m, $L_0=0.25\,$m, $d_i=0.7\,$m, $c_s=13000\,$m$\,$s$^{-1}$ and $\beta_e=20$) yields $f=\gamma_{dr}/(2\pi) \gtrsim 60\,$kHz. This frequency is in remarkable agreement with our experimental results which indicate a transition to the multiple plasmoid regime in the $60-100$ kHz range.

\bibliographystyle{jpp}

\bibliography{references}

\begin{thebibliography}{58}
\expandafter\ifx\csname natexlab\endcsname\relax\def\natexlab#1{#1}\fi
\def\au#1{#1} \def\ed#1{#1} \def\yr#1{#1}\def\at#1{#1}\def\jt#1{\textit{#1}}
  \def\bt#1{#1}\def\bvol#1{\textbf{#1}} \def\vol#1{#1} \def\pg#1{#1}
  \def\publ#1{#1}\def\arxiv#1{#1}\def\org#1{#1}\def\st#1{\textit{#1}}

\bibitem[Antiochos {\em et~al.\/}(2007)Antiochos, DeVore, Karpen \&
  Miki{\'{c}}]{Antiochos2007}
{\sc \au{Antiochos, S.~K.}, \au{DeVore, C.~R.}, \au{Karpen, J.~T.} \&
  \au{Miki{\'{c}}, Z.}} \yr{2007}  \at{{Structure and Dynamics of the Sun's
  Open Magnetic Field}}.  \jt{The Astrophysical Journal}  \bvol{671}~(1),
  \pg{936--946}.

\bibitem[Antiochos {\em et~al.\/}(2011)Antiochos, Miki{\'{c}}, Titov, Lionello
  \& Linker]{Antiochos2011}
{\sc \au{Antiochos, S.~K.}, \au{Miki{\'{c}}, Z.}, \au{Titov, V.~S.},
  \au{Lionello, R.} \& \au{Linker, J.~A.}} \yr{2011}  \at{{A MODEL FOR THE
  SOURCES OF THE SLOW SOLAR WIND}}.  \jt{The Astrophysical Journal}
  \bvol{731}~(2),  \pg{112}.

\bibitem[{Attico} {\em et~al.\/}(2000){Attico}, {Califano} \&
  {Pegoraro}]{Attico_2000}
{\sc \au{{Attico}, N.}, \au{{Califano}, F.} \& \au{{Pegoraro}, F.}} \yr{2000}
  \at{{Fast collisionless reconnection in the whistler frequency range}}.
  \jt{Physics of Plasmas}  \bvol{7}~(6),  \pg{2381--2387}.

\bibitem[Bale {\em et~al.\/}(2019)Bale, Badman, Bonnell, Bowen, Burgess, Case,
  Cattell, Chandran, Chaston, Chen, Drake, de~Wit, Eastwood, Ergun, Farrell,
  Fong, Goetz, Goldstein, Goodrich, Harvey, Horbury, Howes, Kasper, Kellogg,
  Klimchuk, Korreck, Krasnoselskikh, Krucker, Laker, Larson, MacDowall,
  Maksimovic, Malaspina, Martinez-Oliveros, McComas, Meyer-Vernet, Moncuquet,
  Mozer, Phan, Pulupa, Raouafi, Salem, Stansby, Stevens, Szabo, Velli, Woolley
  \& Wygant]{Bale2019}
{\sc \au{Bale, S.~D.}, \au{Badman, S.~T.}, \au{Bonnell, J.~W.}, \au{Bowen,
  T.~A.}, \au{Burgess, D.}, \au{Case, A.~W.}, \au{Cattell, C.~A.},
  \au{Chandran, B. D.G.~G.}, \au{Chaston, C.~C.}, \au{Chen, C. H.K.~K.},
  \au{Drake, J.~F.}, \au{de~Wit, T.~Dudok}, \au{Eastwood, J.~P.}, \au{Ergun,
  R.~E.}, \au{Farrell, W.~M.}, \au{Fong, C.}, \au{Goetz, K.}, \au{Goldstein,
  M.}, \au{Goodrich, K.~A.}, \au{Harvey, P.~R.}, \au{Horbury, T.~S.},
  \au{Howes, G.~G.}, \au{Kasper, J.~C.}, \au{Kellogg, P.~J.}, \au{Klimchuk,
  J.~A.}, \au{Korreck, K.~E.}, \au{Krasnoselskikh, V.~V.}, \au{Krucker, S.},
  \au{Laker, R.}, \au{Larson, D.~E.}, \au{MacDowall, R.~J.}, \au{Maksimovic,
  M.}, \au{Malaspina, D.~M.}, \au{Martinez-Oliveros, J.}, \au{McComas, D.~J.},
  \au{Meyer-Vernet, N.}, \au{Moncuquet, M.}, \au{Mozer, F.~S.}, \au{Phan,
  T.~D.}, \au{Pulupa, M.}, \au{Raouafi, N.~E.}, \au{Salem, C.}, \au{Stansby,
  D.}, \au{Stevens, M.}, \au{Szabo, A.}, \au{Velli, M.}, \au{Woolley, T.} \&
  \au{Wygant, J.~R.}} \yr{2019}  \at{{Highly structured slow solar wind
  emerging from an equatorial coronal hole}}.  \jt{Nature}  \bvol{576}~(7786),
  \pg{237--242}.

\bibitem[Bavassano \& Bruno(1989{\natexlab{{\em a\/}}})]{Bavassano1989}
{\sc \au{Bavassano, B.} \& \au{Bruno, R.}} \yr{1989{\natexlab{{\em a\/}}}}
  \at{{Evidence of local generation of Alfv{\'{e}}nic turbulence in the solar
  wind}}.  \jt{Journal of Geophysical Research}  \bvol{94}~(A9),  \pg{11977}.

\bibitem[Bavassano \& Bruno(1989{\natexlab{{\em b\/}}})]{Bavassano1989a}
{\sc \au{Bavassano, B.} \& \au{Bruno, R.}} \yr{1989{\natexlab{{\em b\/}}}}
  \at{{Large-scale solar wind fluctuations in the inner heliosphere at low
  solar activity}}.  \jt{Journal of Geophysical Research}  \bvol{94}~(A1),
  \pg{168}.

\bibitem[Bavassano {\em et~al.\/}(1982)Bavassano, Dobrowolny, Mariani \&
  Ness]{Bavassano1982}
{\sc \au{Bavassano, B.}, \au{Dobrowolny, M.}, \au{Mariani, F.} \& \au{Ness,
  N.~F.}} \yr{1982}  \at{{Radial evolution of power spectra of interplanetary
  Alfv{\'{e}}nic turbulence}}.  \jt{Journal of Geophysical Research}
  \bvol{87}~(A5),  \pg{3617}.

\bibitem[Bavassano {\em et~al.\/}(1997)Bavassano, Woo \& Bruno]{Bavassano1997}
{\sc \au{Bavassano, Bruno}, \au{Woo, Richard} \& \au{Bruno, Roberto}} \yr{1997}
   \at{{Heliospheric plasma sheet and coronal streamers}}.  \jt{Geophysical
  Research Letters}  \bvol{24}~(13),  \pg{1655--1658}.

\bibitem[Belcher \& Davis(1971)]{Belcher1971}
{\sc \au{Belcher, J.~W.} \& \au{Davis, Leverett}} \yr{1971}
  \at{{Large-amplitude Alfv{\'{e}}n waves in the interplanetary medium, 2}}.
  \jt{Journal of Geophysical Research}  \bvol{76}~(16),  \pg{3534--3563}.

\bibitem[Bhat \& Loureiro(2018)]{Bhat2018}
{\sc \au{Bhat, Pallavi} \& \au{Loureiro, Nuno~F.}} \yr{2018}  \at{{Plasmoid
  instability in the semi-collisional regime}}.  \jt{Journal of Plasma Physics}
   \bvol{84}~(6),  \arxiv{arXiv: 1804.05145}.

\bibitem[Brueckner {\em et~al.\/}(1995)Brueckner, Howard, Koomen, Korendyke,
  Michels, Moses, Socker, Dere, Lamy, Llebaria, Bout, Schwenn, Simnett, Bedford
  \& Eyles]{Brueckner1995}
{\sc \au{Brueckner, G.~E.}, \au{Howard, R.~A.}, \au{Koomen, M.~J.},
  \au{Korendyke, C.~M.}, \au{Michels, D.~J.}, \au{Moses, J.~D.}, \au{Socker,
  D.~G.}, \au{Dere, K.~P.}, \au{Lamy, P.~L.}, \au{Llebaria, A.}, \au{Bout,
  M.~V.}, \au{Schwenn, R.}, \au{Simnett, G.~M.}, \au{Bedford, D.~K.} \&
  \au{Eyles, C.~J.}} \yr{1995}  \at{{The Large Angle Spectroscopic Coronagraph
  (LASCO)}}.  \jt{Solar Physics}  \bvol{162}~(1-2),  \pg{357--402}.

\bibitem[Coleman {\em et~al.\/}(1968)Coleman, J. \& Jr.]{Coleman1968}
{\sc \au{Coleman, Paul~J., Jr.}, \au{J., Paul} \& \au{Jr.}} \yr{1968}
  \at{{Turbulence, Viscosity, and Dissipation in the Solar-Wind Plasma}}.
  \jt{The Astrophysical Journal}  \bvol{153},  \pg{371}.

\bibitem[Cranmer {\em et~al.\/}(2017)Cranmer, Gibson \& Riley]{Cranmer2017}
{\sc \au{Cranmer, Steven~R.}, \au{Gibson, Sarah~E.} \& \au{Riley, Pete}}
  \yr{2017}  \at{{Origins of the Ambient Solar Wind: Implications for Space
  Weather}}.  \jt{Space Science Reviews}  \bvol{212}~(3-4),  \pg{1345--1384}.

\bibitem[Crooker {\em et~al.\/}(2000)Crooker, Shodhan, Gosling, Simmerer,
  Lepping, Steinberg \& Kahler]{Crooker2000}
{\sc \au{Crooker, N.~U.}, \au{Shodhan, S.}, \au{Gosling, J.~T.}, \au{Simmerer,
  J.}, \au{Lepping, R.~P.}, \au{Steinberg, J.~T.} \& \au{Kahler, S.~W.}}
  \yr{2000}  \at{{Density extremes in the solar wind}}.  \jt{Geophysical
  Research Letters}  \bvol{27}~(23),  \pg{3769--3772}.

\bibitem[DeForest {\em et~al.\/}(2018)DeForest, Howard, Velli, Viall \&
  Vourlidas]{DeForest2018}
{\sc \au{DeForest, C.~E.}, \au{Howard, R.~A.}, \au{Velli, M.}, \au{Viall, N.}
  \& \au{Vourlidas, A.}} \yr{2018}  \at{{The Highly Structured Outer Solar
  Corona}}.  \jt{The Astrophysical Journal}  \bvol{862}~(1),  \pg{18}.

\bibitem[{Di Matteo} {\em et~al.\/}(2019){Di Matteo}, Viall, Kepko, Wallace,
  Arge \& MacNeice]{DiMatteo2019}
{\sc \au{{Di Matteo}, S.}, \au{Viall, N.~M.}, \au{Kepko, L.}, \au{Wallace, S.},
  \au{Arge, C.~N.} \& \au{MacNeice, P.}} \yr{2019}  \at{{Helios Observations of
  Quasiperiodic Density Structures in the Slow Solar Wind at 0.3, 0.4, and 0.6
  AU}}.  \jt{Journal of Geophysical Research: Space Physics}  \bvol{124}~(2),
  \pg{837--860}.

\bibitem[Einaudi {\em et~al.\/}(1999)Einaudi, Boncinelli, Dahlburg \&
  Karpen]{Einaudi1999}
{\sc \au{Einaudi, Giorgio}, \au{Boncinelli, Paolo}, \au{Dahlburg, Russell~B.}
  \& \au{Karpen, Judith~T.}} \yr{1999}  \at{{Formation of the slow solar wind
  in a coronal streamer}}.  \jt{Journal of Geophysical Research: Space Physics}
   \bvol{104}~(A1),  \pg{521--534}.

\bibitem[Einaudi {\em et~al.\/}(2001)Einaudi, Chibbaro, Dahlburg \&
  Velli]{Einaudi2001}
{\sc \au{Einaudi, Giorgio}, \au{Chibbaro, Sergio}, \au{Dahlburg, Russell~B.} \&
  \au{Velli, Marco}} \yr{2001}  \at{{Plasmoid Formation and Acceleration in the
  Solar Streamer Belt}}.  \jt{The Astrophysical Journal}  \bvol{547}~(2),
  \pg{1167--1177}.

\bibitem[Endeve {\em et~al.\/}(2004)Endeve, Holzer \& Leer]{Endeve2004}
{\sc \au{Endeve, Eirik}, \au{Holzer, Thomas~E.} \& \au{Leer, Egil}} \yr{2004}
  \at{{Helmet Streamers Gone Unstable: Two-Fluid Magnetohydrodynamic Models of
  the Solar Corona}}.  \jt{The Astrophysical Journal}  \bvol{603}~(1),
  \pg{307--321}.

\bibitem[Endeve {\em et~al.\/}(2003)Endeve, Leer \& Holzer]{Endeve2003}
{\sc \au{Endeve, Eirik}, \au{Leer, Egil} \& \au{Holzer, Thomas~E.}} \yr{2003}
  \at{{Two‐dimensional Magnetohydrodynamic Models of the Solar Corona: Mass
  Loss from the Streamer Belt}}.  \jt{The Astrophysical Journal}
  \bvol{589}~(2),  \pg{1040--1053}.

\bibitem[Endrizzi {\em et~al.\/}(2021)Endrizzi, Egedal, Clark, Flanagan,
  Greess, Milhone, Millet-Ayala, Olson, Peterson, Wallace \&
  Forest]{Endrizzi2021}
{\sc \au{Endrizzi, Douglass}, \au{Egedal, J.}, \au{Clark, M.}, \au{Flanagan,
  K.}, \au{Greess, S.}, \au{Milhone, J.}, \au{Millet-Ayala, A.}, \au{Olson,
  J.}, \au{Peterson, E.~E.}, \au{Wallace, J.} \& \au{Forest, C.~B.}} \yr{2021}
  \at{Laboratory resolved structure of supercritical perpendicular shocks}.
  \jt{Phys. Rev. Lett.}  \bvol{126},  \pg{145001}.

\bibitem[Fisk {\em et~al.\/}(1998)Fisk, Schwadron \& Zurbuchen]{Fisk1998}
{\sc \au{Fisk, L.A.}, \au{Schwadron, N.A.} \& \au{Zurbuchen, T.H.}} \yr{1998}
  \at{{On the Slow Solar Wind}}.  \jt{Space Science Reviews}  \bvol{86}~(1/4),
  \pg{51--60}.

\bibitem[Fisk(2003)]{Fisk2003}
{\sc \au{Fisk, L.~A.}} \yr{2003}  \at{{Acceleration of the solar wind as a
  result of the reconnection of open magnetic flux with coronal loops}}.
  \jt{Journal of Geophysical Research}  \bvol{108}~(A4),  \pg{1157}.

\bibitem[Flanagan {\em et~al.\/}(2020)Flanagan, Milhone, Egedal, Endrizzi,
  Olson, Peterson, Sassella \& Forest]{Flanagan2020}
{\sc \au{Flanagan, K.}, \au{Milhone, J.}, \au{Egedal, J.}, \au{Endrizzi, D.},
  \au{Olson, J.}, \au{Peterson, E.~E.}, \au{Sassella, R.} \& \au{Forest,
  C.~B.}} \yr{2020}  \at{Weakly magnetized, hall dominated plasma couette
  flow}.  \jt{Phys. Rev. Lett.}  \bvol{125},  \pg{135001}.

\bibitem[Forest {\em et~al.\/}(2015)Forest, Flanagan, Brookhart, Clark, Cooper,
  D{\'{e}}sangles, Egedal, Endrizzi, Khalzov, Li, Miesch, Milhone, Nornberg,
  Olson, Peterson, Roesler, Schekochihin, Schmitz, Siller, Spitkovsky, Stemo,
  Wallace, Weisberg \& Zweibel]{Forest2015}
{\sc \au{Forest, C.~B.}, \au{Flanagan, K.}, \au{Brookhart, M.}, \au{Clark, M.},
  \au{Cooper, C.~M.}, \au{D{\'{e}}sangles, V.}, \au{Egedal, J.}, \au{Endrizzi,
  D.}, \au{Khalzov, I.~V.}, \au{Li, H.}, \au{Miesch, M.}, \au{Milhone, J.},
  \au{Nornberg, M.}, \au{Olson, J.}, \au{Peterson, E.}, \au{Roesler, F.},
  \au{Schekochihin, A.}, \au{Schmitz, O.}, \au{Siller, R.}, \au{Spitkovsky,
  A.}, \au{Stemo, A.}, \au{Wallace, J.}, \au{Weisberg, D.} \& \au{Zweibel, E.}}
  \yr{2015}  \at{{The Wisconsin Plasma Astrophysics Laboratory}}.  \jt{Journal
  of Plasma Physics}  \bvol{81}~(5),  \pg{345810501}.

\bibitem[Fu {\em et~al.\/}(2017)Fu, Madjarska, Xia, Li, Huang \&
  Wangguan]{Fu2017}
{\sc \au{Fu, Hui}, \au{Madjarska, Maria~S.}, \au{Xia, LiDong}, \au{Li, Bo},
  \au{Huang, ZhengHua} \& \au{Wangguan, Zhipeng}} \yr{2017}  \at{{Charge States
  and FIP Bias of the Solar Wind from Coronal Holes, Active Regions, and Quiet
  Sun}}.  \jt{The Astrophysical Journal}  \bvol{836}~(2),  \pg{169}.

\bibitem[Hare {\em et~al.\/}(2017)Hare, Suttle, Lebedev, Loureiro, Ciardi,
  Burdiak, Chittenden, Clayson, Garcia, Niasse, Robinson, Smith, Stuart,
  Suzuki-Vidal, Swadling, Ma, Wu \& Yang]{hare_2017}
{\sc \au{Hare, J.~D.}, \au{Suttle, L.}, \au{Lebedev, S.~V.}, \au{Loureiro,
  N.~F.}, \au{Ciardi, A.}, \au{Burdiak, G.~C.}, \au{Chittenden, J.~P.},
  \au{Clayson, T.}, \au{Garcia, C.}, \au{Niasse, N.}, \au{Robinson, T.},
  \au{Smith, R.~A.}, \au{Stuart, N.}, \au{Suzuki-Vidal, F.}, \au{Swadling,
  G.~F.}, \au{Ma, J.}, \au{Wu, J.} \& \au{Yang, Q.}} \yr{2017}  \at{Anomalous
  heating and plasmoid formation in a driven magnetic reconnection experiment}.
   \jt{Physical Review Letters}  \bvol{118},  \pg{085001}.

\bibitem[Higginson \& Lynch(2018)]{Higginson2018}
{\sc \au{Higginson, A.~K.} \& \au{Lynch, B.~J.}} \yr{2018}  \at{{Structured
  Slow Solar Wind Variability: Streamer-blob Flux Ropes and Torsional
  Alfv{\'{e}}n Waves}}.  \jt{The Astrophysical Journal}  \bvol{859}~(1),
  \pg{6}.

\bibitem[Kepko \& Spence(2003)]{Kepko2003}
{\sc \au{Kepko, L.} \& \au{Spence, H.~E.}} \yr{2003}  \at{{Observations of
  discrete, global magnetospheric oscillations directly driven by solar wind
  density variations}}.  \jt{Journal of Geophysical Research}  \bvol{108}~(A6),
   \pg{1257}.

\bibitem[Kepko {\em et~al.\/}(2002)Kepko, Spence \& Singer]{Kepko2002}
{\sc \au{Kepko, L.}, \au{Spence, H.~E.} \& \au{Singer, H.~J.}} \yr{2002}
  \at{{ULF waves in the solar wind as direct drivers of magnetospheric
  pulsations}}.  \jt{Geophysical Research Letters}  \bvol{29}~(8),
  \pg{39--1--39--4}.

\bibitem[Kepko {\em et~al.\/}(2016)Kepko, Viall, Antiochos, Lepri, Kasper \&
  Weberg]{Kepko2016}
{\sc \au{Kepko, L.}, \au{Viall, N.~M.}, \au{Antiochos, S.~K.}, \au{Lepri,
  S.~T.}, \au{Kasper, J.~C.} \& \au{Weberg, M.}} \yr{2016}  \at{{Implications
  of L1 observations for slow solar wind formation by solar reconnection}}.
  \jt{Geophysical Research Letters}  \bvol{43}~(9),  \pg{4089--4097}.

\bibitem[K{\"o}hnlein(1996)]{Kohnlein1996}
{\sc \au{K{\"o}hnlein, W.}} \yr{1996}  \at{{Radial dependence of solar wind
  parameters in the ecliptic (1.1 {$R_\odot$} - 61{AU})}}.  \jt{Solar Physics}
  \bvol{169}~(1),  \pg{209--213}.

\bibitem[Lapenta \& Knoll(2005)]{Lapenta2005}
{\sc \au{Lapenta, Giovanni} \& \au{Knoll, D.~A.}} \yr{2005}  \at{{Effect of a
  Converging Flow at the Streamer Cusp on the Genesis of the Slow Solar Wind}}.
   \jt{The Astrophysical Journal}  \bvol{624}~(2),  \pg{1049--1056}.

\bibitem[Lavraud {\em et~al.\/}(2020)Lavraud, Fargette, R{\'{e}}ville, Szabo,
  Huang, Rouillard, Viall, Phan, Kasper, Bale, Berthomier, Bonnell, Case,
  de~Wit, Eastwood, G{\'{e}}not, Goetz, Griton, Halekas, Harvey, Kieokaew,
  Klein, Korreck, Kouloumvakos, Larson, Lavarra, Livi, Louarn, MacDowall,
  Maksimovic, Malaspina, Nieves-Chinchilla, Pinto, Poirier, Pulupa, Raouafi,
  Stevens, Toledo-Redondo \& Whittlesey]{Lavraud_2020}
{\sc \au{Lavraud, B.}, \au{Fargette, N.}, \au{R{\'{e}}ville, V.}, \au{Szabo,
  A.}, \au{Huang, J.}, \au{Rouillard, A.~P.}, \au{Viall, N.}, \au{Phan, T.~D.},
  \au{Kasper, J.~C.}, \au{Bale, S.~D.}, \au{Berthomier, M.}, \au{Bonnell,
  J.~W.}, \au{Case, A.~W.}, \au{de~Wit, T.~Dudok}, \au{Eastwood, J.~P.},
  \au{G{\'{e}}not, V.}, \au{Goetz, K.}, \au{Griton, L.~S.}, \au{Halekas, J.~S},
  \au{Harvey, P.}, \au{Kieokaew, R.}, \au{Klein, K.~G.}, \au{Korreck, K.~E.},
  \au{Kouloumvakos, A.}, \au{Larson, D.~E.}, \au{Lavarra, M.}, \au{Livi, R.},
  \au{Louarn, P.}, \au{MacDowall, R.~J.}, \au{Maksimovic, M.}, \au{Malaspina,
  D.}, \au{Nieves-Chinchilla, T.}, \au{Pinto, R.~F.}, \au{Poirier, N.},
  \au{Pulupa, M.}, \au{Raouafi, N.~E.}, \au{Stevens, M.~L.},
  \au{Toledo-Redondo, S.} \& \au{Whittlesey, P.~L.}} \yr{2020}  \at{The
  heliospheric current sheet and plasma sheet during parker solar probe's first
  orbit}.  \jt{The Astrophysical Journal}  \bvol{894}~(2),  \pg{L19}.

\bibitem[{Loureiro} {\em et~al.\/}(2012){Loureiro}, {Samtaney}, {Schekochihin}
  \& {Uzdensky}]{loureiro_2012}
{\sc \au{{Loureiro}, N.~F.}, \au{{Samtaney}, R.}, \au{{Schekochihin}, A.~A.} \&
  \au{{Uzdensky}, D.~A.}} \yr{2012}  \at{{Magnetic reconnection and stochastic
  plasmoid chains in high-Lundquist-number plasmas}}.  \jt{Physics of Plasmas}
  \bvol{19}~(4),  \pg{042303--042303},  \arxiv{arXiv: 1108.4040}.

\bibitem[Luttrell \& Richter(1987)]{Luttrell1987}
{\sc \au{Luttrell, A.~H.} \& \au{Richter, A.~K.}} \yr{1987}  \at{{The Role of
  Alfvenic Fluctuations in MHD Turbulence Evolution between 0.3 and 1.0 AU}}.
  \jt{Sixth International Solar Wind Conference, Proceedings of the conference
  held 23-28 August, 1987 at YMCA of the Rockies, Estes Park, Colorado. Edited
  by V.J. Pizzo, T. Holzer, and D.G. Sime. NCAR Technical Note
  NCAR/TN-306+Proc, Volume 2, 1987., p.335}  \pg{p. 335}.

\bibitem[Marsch \& Tu(1990{\natexlab{{\em a\/}}})]{Marsch1990a}
{\sc \au{Marsch, E.} \& \au{Tu, C.-Y.}} \yr{1990{\natexlab{{\em a\/}}}}
  \at{{On the radial evolution of MHD turbulence in the inner heliosphere}}.
  \jt{Journal of Geophysical Research}  \bvol{95}~(A6),  \pg{8211}.

\bibitem[Marsch \& Tu(1990{\natexlab{{\em b\/}}})]{Marsch1990b}
{\sc \au{Marsch, E.} \& \au{Tu, C.-Y.}} \yr{1990{\natexlab{{\em b\/}}}}
  \at{{Spectral and spatial evolution of compressible turbulence in the inner
  solar wind}}.  \jt{Journal of Geophysical Research}  \bvol{95}~(A8),
  \pg{11945}.

\bibitem[Neugebauer {\em et~al.\/}(2016)Neugebauer, Reisenfeld \&
  Richardson]{Neugebauer2016}
{\sc \au{Neugebauer, Marcia}, \au{Reisenfeld, Daniel} \& \au{Richardson,
  Ian~G.}} \yr{2016}  \at{{Comparison of algorithms for determination of solar
  wind regimes}}.  \jt{Journal of Geophysical Research: Space Physics}
  \bvol{121}~(9),  \pg{8215--8227}.

\bibitem[Neugebauer \& Snyder(1962)]{Neugebauer1962}
{\sc \au{Neugebauer, Marcia} \& \au{Snyder, Conway~W}} \yr{1962}  \at{{Solar
  Plasma Experiment}}.  \jt{Science}  \bvol{138}~(3545),  \pg{1095--1097}.

\bibitem[Olson {\em et~al.\/}(2016)Olson, Egedal, Greess, Myers, Clark,
  Endrizzi, Flanagan, Milhone, Peterson, Wallace, Weisberg \&
  Forest]{Olson2016}
{\sc \au{Olson, J.}, \au{Egedal, J.}, \au{Greess, S.}, \au{Myers, R.},
  \au{Clark, M.}, \au{Endrizzi, D.}, \au{Flanagan, K.}, \au{Milhone, J.},
  \au{Peterson, E.}, \au{Wallace, J.}, \au{Weisberg, D.} \& \au{Forest, C.B.}}
  \yr{2016}  \at{{Experimental Demonstration of the Collisionless Plasmoid
  Instability below the Ion Kinetic Scale during Magnetic Reconnection}}.
  \jt{Physical Review Letters}  \bvol{116}~(25),  \pg{1--5}.

\bibitem[Parker(1958)]{Parker1958a}
{\sc \au{Parker, Eugene~N.}} \yr{1958}  \at{{Dynamics of the Interplanetary Gas
  and Magnetic Fields.}}  \jt{The Astrophysical Journal}  \bvol{128},
  \pg{664--676}.

\bibitem[Peterson {\em et~al.\/}(2019)Peterson, Endrizzi, Beidler, Bunkers,
  Clark, Egedal, Flanagan, McCollam, Milhone, Olson, Sovinec, Waleffe, Wallace
  \& Forest]{Peterson2019}
{\sc \au{Peterson, Ethan~E.}, \au{Endrizzi, Douglass~A.}, \au{Beidler,
  Matthew}, \au{Bunkers, Kyle~J.}, \au{Clark, Michael}, \au{Egedal, Jan},
  \au{Flanagan, Ken}, \au{McCollam, Karsten~J.}, \au{Milhone, Jason},
  \au{Olson, Joseph}, \au{Sovinec, Carl~R.}, \au{Waleffe, Roger}, \au{Wallace,
  John} \& \au{Forest, Cary~B.}} \yr{2019}  \at{{A laboratory model for the
  Parker spiral and magnetized stellar winds}}.  \jt{Nature Physics}
  \pg{p.~1}.

\bibitem[Rouillard {\em et~al.\/}(2011)Rouillard, Sheeley, Cooper, Davies,
  Lavraud, Kilpua, Skoug, Steinberg, Szabo, Opitz \& Sauvaud]{Rouillard2011}
{\sc \au{Rouillard, A.~P.}, \au{Sheeley, N.~R.}, \au{Cooper, T.~J.},
  \au{Davies, J.~A.}, \au{Lavraud, B.}, \au{Kilpua, E. K.~J.}, \au{Skoug,
  R.~M.}, \au{Steinberg, J.~T.}, \au{Szabo, A.}, \au{Opitz, A.} \& \au{Sauvaud,
  J.-A.}} \yr{2011}  \at{{THE SOLAR ORIGIN OF SMALL INTERPLANETARY
  TRANSIENTS}}.  \jt{The Astrophysical Journal}  \bvol{734}~(1),  \pg{7}.

\bibitem[Sanchez-Diaz {\em et~al.\/}(2019)Sanchez-Diaz, Rouillard, Lavraud,
  Kilpua \& Davies]{Sanchez-Diaz2019}
{\sc \au{Sanchez-Diaz, E.}, \au{Rouillard, A.~P.}, \au{Lavraud, B.},
  \au{Kilpua, E.} \& \au{Davies, J.~A.}} \yr{2019}  \at{{In Situ Measurements
  of the Variable Slow Solar Wind near Sector Boundaries}}.  \jt{The
  Astrophysical Journal}  \bvol{882}~(1),  \pg{51},  \arxiv{arXiv: 1911.09683}.

\bibitem[Sheeley {\em et~al.\/}(2009)Sheeley, Lee, Casto, Wang \&
  Rich]{Sheeley2009}
{\sc \au{Sheeley, N.~R.}, \au{Lee, D. D.-H.}, \au{Casto, K.~P.}, \au{Wang,
  Y.-M.} \& \au{Rich, N.~B.}} \yr{2009}  \at{{THE STRUCTURE OF STREAMER
  BLOBS}}.  \jt{The Astrophysical Journal}  \bvol{694}~(2),  \pg{1471--1480}.

\bibitem[{Sheeley, Jr.} {\em et~al.\/}(1997){Sheeley, Jr.}, Wang, Hawley,
  Brueckner, Dere, Howard, Koomen, Korendyke, Michels, Paswaters, Socker, {St.
  Cyr}, Wang, Lamy, Llebaria, Schwenn, Simnett, Plunkett \&
  Biesecker]{SheeleyJr.1997}
{\sc \au{{Sheeley, Jr.}, N.~R.}, \au{Wang, Y.‐M.}, \au{Hawley, S.~H.},
  \au{Brueckner, G.~E.}, \au{Dere, K.~P.}, \au{Howard, R.~A.}, \au{Koomen,
  M.~J.}, \au{Korendyke, C.~M.}, \au{Michels, D.~J.}, \au{Paswaters, S.~E.},
  \au{Socker, D.~G.}, \au{{St. Cyr}, O.~C.}, \au{Wang, D.}, \au{Lamy, P.~L.},
  \au{Llebaria, A.}, \au{Schwenn, R.}, \au{Simnett, G.~M.}, \au{Plunkett, S.}
  \& \au{Biesecker, D.~A.}} \yr{1997}  \at{{Measurements of Flow Speeds in the
  Corona Between 2 and 30{$R_\odot$} }}.  \jt{The Astrophysical Journal}
  \bvol{484}~(1),  \pg{472--478}.

\bibitem[Sovinec {\em et~al.\/}(2004)Sovinec, Glasser, Gianakon, Barnes, Nebel,
  Kruger, Schnack, Plimpton, Tarditi \& Chu]{Sovinec2004}
{\sc \au{Sovinec, C.R.}, \au{Glasser, A.H.}, \au{Gianakon, T.A.}, \au{Barnes,
  D.C.}, \au{Nebel, R.A.}, \au{Kruger, S.E.}, \au{Schnack, D.D.}, \au{Plimpton,
  S.J.}, \au{Tarditi, A.} \& \au{Chu, M.S.}} \yr{2004}  \at{{Nonlinear
  magnetohydrodynamics simulation using high-order finite elements}}.
  \jt{Journal of Computational Physics}  \bvol{195}~(1),  \pg{355--386}.

\bibitem[Sovinec \& King(2010)]{Sovinec2010}
{\sc \au{Sovinec, C.R.} \& \au{King, J.R.}} \yr{2010}  \at{{Analysis of a mixed
  semi-implicit/implicit algorithm for low-frequency two-fluid plasma
  modeling}}.  \jt{Journal of Computational Physics}  \bvol{229}~(16),
  \pg{5803--5819}.

\bibitem[Stephenson \& Walker(2002)]{Stephenson2002}
{\sc \au{Stephenson, J. A.~E.} \& \au{Walker, A. D.~M.}} \yr{2002}  \at{{HF
  radar observations of Pc5 ULF pulsations driven by the solar wind}}.
  \jt{Geophysical Research Letters}  \bvol{29}~(9),  \pg{8--1--8--4}.

\bibitem[Uzdensky \& Loureiro(2016)]{Uzdensky2016}
{\sc \au{Uzdensky, D.~A.} \& \au{Loureiro, N.~F.}} \yr{2016}  \at{{Magnetic
  Reconnection Onset via Disruption of a Forming Current Sheet by the Tearing
  Instability}}.  \jt{Physical Review Letters}  \bvol{116}~(10),  \arxiv{arXiv:
  1411.4295}.

\bibitem[Uzdensky {\em et~al.\/}(2010)Uzdensky, Loureiro \&
  Schekochihin]{uzdensky_2010}
{\sc \au{Uzdensky, D.~A.}, \au{Loureiro, N.~F.} \& \au{Schekochihin, A.~A.}}
  \yr{2010}  \at{Fast magnetic reconnection in the plasmoid-dominated regime}.
  \jt{Physical Review Letters}  \bvol{105},  \pg{235002}.

\bibitem[Viall {\em et~al.\/}(2008)Viall, Kepko \& Spence]{Viall2008}
{\sc \au{Viall, N.~M.}, \au{Kepko, L.} \& \au{Spence, H.~E.}} \yr{2008}
  \at{{Inherent length-scales of periodic solar wind number density
  structures}}.  \jt{Journal of Geophysical Research: Space Physics}
  \bvol{113}~(A7),  \pg{n/a--n/a}.

\bibitem[Viall \& Vourlidas(2015)]{Viall2015}
{\sc \au{Viall, Nicholeen~M.} \& \au{Vourlidas, Angelos}} \yr{2015}
  \at{{PERIODIC DENSITY STRUCTURES AND THE ORIGIN OF THE SLOW SOLAR WIND}}.
  \jt{The Astrophysical Journal}  \bvol{807}~(2),  \pg{176}.

\bibitem[Wang {\em et~al.\/}(1997)Wang, {Sheeley, Jr.}, Howard, Kraemer, Rich,
  Andrews, Brueckner, Dere, Koomen, Korendyke, Michels, Moses, Paswaters,
  Socker, Wang, Lamy, Llebaria, Vibert, Schwenn \& Simnett]{Wang1997}
{\sc \au{Wang, Y.‐M.}, \au{{Sheeley, Jr.}, N.~R.}, \au{Howard, R.~A.},
  \au{Kraemer, J.~R.}, \au{Rich, N.~B.}, \au{Andrews, M.~D.}, \au{Brueckner,
  G.~E.}, \au{Dere, K.~P.}, \au{Koomen, M.~J.}, \au{Korendyke, C.~M.},
  \au{Michels, D.~J.}, \au{Moses, J.~D.}, \au{Paswaters, S.~E.}, \au{Socker,
  D.~G.}, \au{Wang, D.}, \au{Lamy, P.~L.}, \au{Llebaria, A.}, \au{Vibert, D.},
  \au{Schwenn, R.} \& \au{Simnett, G.~M.}} \yr{1997}  \at{{Origin and Evolution
  of Coronal Streamer Structure During the 1996 Minimum Activity Phase}}.
  \jt{The Astrophysical Journal}  \bvol{485}~(2),  \pg{875--889}.

\bibitem[Wang \& Hess(2018)]{Wang2018}
{\sc \au{Wang, Y.-M.} \& \au{Hess, P.}} \yr{2018}  \at{{Gradual Streamer
  Expansions and the Relationship between Blobs and Inflows}}.  \jt{The
  Astrophysical Journal}  \bvol{859}~(2),  \pg{135}.

\bibitem[Wang {\em et~al.\/}(1998)Wang, {Sheeley, Jr.}, Walters, Brueckner,
  Howard, Michels, Lamy, Schwenn \& Simnett]{Wang1998}
{\sc \au{Wang, Y.-M.}, \au{{Sheeley, Jr.}, N.~R.}, \au{Walters, J.~H.},
  \au{Brueckner, G.~E.}, \au{Howard, R.~A.}, \au{Michels, D.~J.}, \au{Lamy,
  P.~L.}, \au{Schwenn, R.} \& \au{Simnett, G.~M.}} \yr{1998}  \at{{Origin of
  Streamer Material in the Outer Corona}}.  \jt{The Astrophysical Journal}
  \bvol{498}~(2),  \pg{L165--L168}.

\bibitem[Wu {\em et~al.\/}(2000)Wu, Wang, Plunkett \& Michels]{Wu2000}
{\sc \au{Wu, S.~T.}, \au{Wang, A.~H.}, \au{Plunkett, S.~P.} \& \au{Michels,
  D.~J.}} \yr{2000}  \at{{Evolution of Global‐Scale Coronal Magnetic Field
  due to Magnetic Reconnection: The Formation of the Observed Blob Motion in
  the Coronal Streamer Belt}}.  \jt{The Astrophysical Journal}  \bvol{545}~(2),
   \pg{1101--1115}.

\end{thebibliography}

\end{document}